\documentclass[twocolumn,aps,showpacs,superscriptaddress]{revtex4}

\usepackage{amsmath,amssymb,amsmath}
\usepackage{graphicx}
\usepackage{dcolumn}
\usepackage{bm}
\usepackage{multirow}

\begin{document}

\title{Magnetic trapping of ultracold Rydberg atoms in low angular momentum states}

\author{Michael Mayle}
\affiliation{Theoretische Chemie, Universit\"at Heidelberg, D-69120 Heidelberg,
Germany}

\author{Igor Lesanovsky}
\affiliation{Midlands Ultracold Atom
Research Centre - MUARC, The University of Nottingham,
School of Physics and Astronomy, Nottingham, United Kingdom}

\author{Peter Schmelcher}
\affiliation{Theoretische Chemie, Universit\"at Heidelberg, D-69120 Heidelberg,
Germany}
\affiliation{Physikalisches Institut, Universit\"at Heidelberg, D-69120 Heidelberg, Germany}

\date{\today}

\begin{abstract}
We theoretically investigate the quantum properties of
$nS$, $nP$, and $nD$
Rydberg atoms in a magnetic Ioffe-Pritchard trap. In particular, 
it is demonstrated that the two-body character
of Rydberg atoms significantly alters the trapping properties
opposed to point-like particles with identical magnetic moment.
Approximate analytical expressions describing the resulting Rydberg trapping
potentials are derived
and their validity is confirmed for experimentally relevant field strengths
by comparisons to numerical solutions of the underlying Schr\"odinger
equation.
In addition to the electronic properties, the center of mass
dynamics of trapped Rydberg atoms is
studied. In particular, we analyze 
the influence of a short-time Rydberg excitation,
as required by certain quantum-information protocols,
on the center of mass dynamics of trapped ground state atoms.
A corresponding heating rate is derived and the
implications for the purity of the density matrix of an encoded qubit are
investigated.
\end{abstract}

\pacs{32.10.Ee, 
32.80.Ee, 
32.60.+i, 
37.10.Gh  
}

\maketitle

\section{Introduction}
During the past decade, powerful cooling techniques enabled remarkable
experiments with ultracold atomic gases revealing a plethora of intriguing phenomena.
Among the many fascinating systems
are Rydberg atoms possessing extraordinary properties \cite{gallagher94}.
Because of the large displacement of the valence electron and the ionic core, they
are highly polarizable and, therefore, experience a strong dipole-dipole
interaction amongst each other.
In ultracold gases,
the latter has been shown
theoretically \cite{PhysRevLett.85.2208,PhysRevLett.87.037901} and
experimentally \cite{tong:063001,singer:163001,liebisch:253002,vogt:073002,
ditzhuijzen:243201} to entail a blockade mechanism
thereby effectuating a collective excitation process of Rydberg atoms
\cite{heidemann:163601,reetz-lamour:253001,johnson:113003}.
Moreover, two recent experiments demonstrated the blockade between
two single atoms a few $\mu$m apart \cite{Urban2009,Gaetan2009}.
The dipole-dipole interaction
renders Rydberg atoms also promising candidates for the implementation
of protocols realizing two-qubit
quantum gates \cite{PhysRevLett.85.2208,PhysRevLett.87.037901}.
A prerequisite for the latter is, however, the availability of suitable
environments enabling the controlled manipulation of single Rydberg atoms
and preventing the dephasing of ground and Rydberg state.

Several works have focused on the issue of trapping Rydberg
atoms based on electric \cite{hyafil:103001,hogan:043001}, optical \cite{dutta00},
or strong magnetic fields \cite{choi:243001,choi:253005}.
Being omnipresent in experiments dealing with ultracold atoms,
inhomogeneous magnetic fields seem predestined for trapping Rydberg atoms
(even a two-dimensional permanent magnetic lattice of Ioffe-Pritchard
microtraps for ultracold atoms has been realized experimentally
\cite{gerritsma:033408,Whitlock2009}).
Similar to ground state atoms, the magnetic trapping of Rydberg atoms
originates from the interaction of its magnetic moment with the
magnetic field.
In particular, this allows utilizing trap geometries which are
well-known from ground state atoms.
In this spirit,
theoretical studies recently demonstrated that Rydberg atoms can be tightly
confined in a magnetic Ioffe-Pritchard (IP) trap
\cite{hezel:223001,hezel:053417}
and that one-dimensional Rydberg gases can be created and stabilized
by means of an additional electric field \cite{mayle:113004}.
However, the trapping mechanism relies in these studies
on high angular momentum electronic states
that have not been realized yet in experiments with ultracold atoms.
In a very recent work, the authors expanded the former studies
to low angular momentum $nS_{1/2}$ states and showed that the composite
nature of Rydberg atoms, i.e., the fact that they consist of an outer electron far
away from a compact ionic core, significantly alters their trapping properties
opposed to point-like particles with the same magnetic moment
\cite{mayle:041403}.
Furthermore, it has been demonstrated how the specific features of the
Rydberg trapping potential can be probed by means of ground state atoms
that are off-resonantly coupled to the Rydberg state via a two photon laser
transition.
In the present work, we provide a detailed derivation and discussion
of the Rydberg energy surfaces presented in Ref.\ \cite{mayle:041403}.
Moreover, the trapping potentials arising for
the $nS$, $nP$, and $nD$ states of $^{87}$Rb are explored.
As we are going to show, 
they possess a reduced azimuthal symmetry and a 
finite trap depth, which can be a few vibrational quanta only or less.
Choosing the magnetic field parameters appropriately, on the other hand,
trapping can be achieved with trap depths in the micro-Kelvin regime.
Implications for quantum information protocols involving magnetically
trapped Rydberg atoms are discussed.

In detail we proceed as follows.
Section \ref{sec:hamiltonian} contains a derivation of our working
Hamiltonian for low angular momentum Rydberg atoms in a Ioffe-Pritchard
trap which is solved by means of a hybrid computational approach
employing basis-set and discretization techniques.
Section \ref{sec:approximation} then introduces reasonable
approximations which allow us to gain analytical solutions
for the stationary Schr\"odinger equation and hence for
the trapping potentials.
In Sec.~\ref{sec:surfaces} we analyze the resulting energy surfaces which
serve as a potential for the center of mass motion of the Rydberg atom.
The range of validity of our analytical approach is discussed.
Section \ref{sec:cmwave} is dedicated to the c.m.\ dynamics
within the adiabatic potential surfaces.
The question of how the c.m.\ state of a ground state atom is altered due
to its short-time excitation to a Rydberg state is illuminated in
Sec.~\ref{sec:heating}. A heating rate associated with this process is
derived.
In Sec.~\ref{sec:dephasing}, the effect of the same process on the purity 
of the density matrix of a qubit which is encoded in the hyperfine states 
of a ground state atom is discussed.

\section{Hamiltonian}
\label{sec:hamiltonian}
Along the lines of Ref.~\cite{hezel:053417} we model the mutual interaction
of the highly excited valence electron and the remaining closed-shell
ionic core of an alkali Rydberg atom by an effective potential which is assumed
to depend only on the distance of the two particles.
In our previous works \cite{lesanovsky:053001,hezel:053417,hezel:223001,mayle:113004},
this potential could be considered to be purely Coulombic since
solely circular states with maximum electronic angular momentum
were investigated.
The low angular momentum states of alkali atoms, on the other hand,
significantly differ from the hydrogenic ones because of the finite size
and the electronic structure of the ionic core.
However, the resulting core penetration, scattering, and polarization effects
can be accounted for by employing a model potential of the form 
\begin{equation}
 V(r)\equiv V_l(r)=-\frac{Z_l(r)}{r}-\frac{\alpha_c}{2r^4}\big[1-e^{-(r/r_c)^6}\big],
\end{equation}
where $\alpha_c$ is the static dipole polarizability of the positive-ion core while
the radial charge $Z_l(r)$ is given by
\begin{equation}
 Z_l(r)=1+(z-1)e^{-a_1r}-r(a_3+a_4r)e^{-a_2r},
\end{equation}
where $z$ is the nuclear charge of the neutral atom and $r_c$ is the cutoff radius
introduced to truncate the unphysical short-range behavior of the polarization
potential near the origin \cite{PhysRevA.49.982}.
Note that $V_l(r)$ depends on the orbital angular momentum $l$
via its parameters, i.e., $a_i\equiv a_i(l)$ and $r_c\equiv r_c(l)$.
The resulting binding energies are related to the effective quantum number $n^*$
and the quantum defect $\delta$ by
$W=-\frac{1}{2{n^*}^2}=-\frac{1}{2(n-\delta)^2}$
\cite{gallagher94};
unless stated otherwise, all quantities are given in atomic units.

The coupling of the charged particles to the external magnetic
field is introduced via the minimal coupling,
$\mathbf p_i\rightarrow \mathbf p_i-q_i\mathbf A(\mathbf r_i)$,
with $i\in\{e,c\}$ denoting the valence electron and the remaining ionic
core of a Rydberg atom, respectively; $q_i$ is the charge of the
$i$-th particle and $\mathbf{A(x)}$ is the vector potential belonging
to the magnetic field $\mathbf{B(x)}$.
Including the coupling of the magnetic moments to the external field
($\boldsymbol{\mu}_e$ and $\boldsymbol{\mu}_c$ originate from the
electronic and nuclear spins, respectively), our initial Hamiltonian
in the laboratory frame reads
(employing $q_e=-1$, $q_c=1$)
\begin{align}
 H={}&\frac{1}{2}[\mathbf{p}_e+\mathbf{A}(\mathbf{r}_e)]^2
  +\frac{1}{2M}[\mathbf{p}_c-\mathbf{A}(\mathbf{r}_c)]^2\nonumber\\
& -\boldsymbol{\mu}_e\cdot\mathbf{B}(\mathbf{r}_e)
  -\boldsymbol{\mu}_c\cdot\mathbf{B}(\mathbf{r}_c)
+V_l(r)+V_{so}(\mathbf{L}_r,\mathbf{S})
\label{eq:ham_first}
\end{align}
with $r=|\mathbf{r}_e-\mathbf{r}_c|$ and $M$ being the mass
of the ionic core.
In contrast to our previous studies 
\cite{lesanovsky:053001,hezel:223001,hezel:053417,mayle:113004},
one has to take into account
the fine structure of the atomic energy levels:
For the magnetic field strengths investigated in this work,
the spin-orbit interaction will lead to splittings larger
than any Zeeman splitting encountered; it is given by
\begin{equation}\label{eq:surfvso}
 V_{so}(\mathbf{L}_r,\mathbf{S})=\frac{\alpha^2}{2}
 \big[1-\frac{\alpha^2}{2}V_l(r)\big]^{-2}\frac{1}{r}
\frac{\mathrm d V_l(r)}{\mathrm d r}\mathbf L_r\cdot\mathbf S
\end{equation}
where $\mathbf L_r$ and $\mathbf S$ denote the angular momentum and spin
of the valence electron, respectively. The term $[1-\alpha^2V_l(r)/2]^{-2}$
has been introduced to regularize the nonphysical divergence near the
origin \cite{condon35}.
As usual, the field-free electronic eigenstates are labeled
by the total electronic angular momentum $\mathbf{J=L}_r+\mathbf S$.
We remark that the model potential $V_l(r)$ has been developed ignoring 
the fine structure; let us therefore briefly comment on the accuracy 
of Eq.~(\ref{eq:surfvso}) in reproducing the fine structure intervals.
For the $40P$ state of rubidium, our approach yields a good quantitative
agreement with the experimentally determined fine structure splitting, 
showing a deviation of less than one percent. This accuracy decreases 
for higher angular momenta; for the $40D$ state nevertheless a qualitative
 agreement is found.

The magnetic field configuration of the Ioffe-Pritchard trap is given 
by a two-dimensional quadrupole field in the $x_1,x_2$-plane together 
with a perpendicular offset (Ioffe-) field in the $x_3$-direction. It 
can be created by several means. The ``traditional'' macroscopic 
realization uses four parallel current carrying Ioffe bars which generate
 the two-dimensional quadrupole field. Encompassing Helmholtz 
coils create the additional constant field \cite{PhysRevLett.51.1336}. 
More recent implementations are for example the \textsc{quic} 
\cite{PhysRevA.58.R2664} and the clover-leaf configuration 
\cite{PhysRevLett.77.416}. On a microscopic scale, the 
Ioffe-Pritchard trap has been implemented on atom chips by a Z-shaped 
wire \cite{fortagh:235}. The IP configuration can be parametrized as
$\mathbf{B}(\mathbf{x})=\mathbf B_c+\mathbf{B}_l
(\mathbf{x})$
with
$\mathbf B_c=B\mathbf e_3$ and $\mathbf{B}_l(\mathbf{x})=
G\left[x_1\mathbf{e}_1-x_2\mathbf{e}_2\right]$.
The corresponding vector potential reads
$\mathbf{A}(\mathbf{x})=
\mathbf{A}_c(\mathbf{x})+\mathbf{A}_l(\mathbf{x})$
with
$\mathbf{A}_c(\mathbf{x})=
\frac{B}{2}\left[x_1\mathbf{e}_2-x_2\mathbf{e}_1\right]$
and $\mathbf{A}_l(\mathbf{x})=Gx_1x_2\mathbf{e}_3$, where $B$
and $G$ are the Ioffe field strength and the gradient, respectively.
The quadratic term $\mathbf B_q\propto (x_3^2-\rho^2/2)\mathbf e_3$
that usually arises for a IP 
configuration can be exactly zeroed by geometry, which we are considering 
in the following. In actual experimental setups, $\mathbf B_q$
provides a weak confinement also in the $x_3$-direction.
Omitting $\mathbf B_q$, the magnitude of the magnetic field at a 
certain position $\mathbf x$ in 
space is given by $|\mathbf B(\mathbf x)|=\sqrt{B^2+G^2\rho^2}$,
which yields a linear asymptote $|\mathbf B(\mathbf x)|\rightarrow G\rho$ for 
large coordinates ($\rho=\sqrt{x_1^2+x_2^2}\gg B/G$) and a harmonic behavior 
$|\mathbf B(\mathbf x)|\approx B+\frac{1}{2}G^2\rho^2$ close to 
the origin ($\rho\ll B/G$).

After introducing relative and c.m.\ coordinates ($\mathbf{r}$ and
$\mathbf{R}$)
\footnote{we approximate $M+m_e\approx M$,
$m=m_eM/(m_e+M)\approx m_e=1\,$a.u., and $m^{-1}+M^{-1}\approx m^{-1}$.}
and employing the unitary transformation
$U=\exp\left[-\frac{i}{2}(\mathbf{B}_c\times \mathbf{r}) \cdot
\mathbf{R}\right]$, the Hamiltonian describing the
Rydberg atom becomes
\begin{align}
\label{eq:hamfinaluni}
U HU^\dagger={}&H_A+\frac{\mathbf{P}^2}{2M}
+\tfrac{1}{2}[\mathbf L_r+2\mathbf S]\cdot\mathbf B_c
-\boldsymbol{\mu}_e\cdot\mathbf{B}_l(\mathbf{R+r})
\nonumber\\
&-\boldsymbol{\mu}_c\cdot\mathbf{B}(\mathbf{R})
+\mathbf{A}_l(\mathbf{R+r})\cdot\mathbf{p}
+\tfrac{1}{2}\mathbf{A}_c(\mathbf{r})^2+H_\mathrm{corr}\,.
\end{align}
Here, $H_A=\mathbf{p}^2/2+V_l(r)+V_{so}(\mathbf{L}_r,\mathbf{S})$
is the Hamiltonian of an alkali atom possessing the energies
$E_{nlj}^{el}=-\frac{1}{2}(n-\delta_{nlj})^{-2}$.
$H_\mathrm{corr}=
\frac{1}{2}\mathbf A_l(\mathbf{R+r})^2
+\frac{1}{M}\mathbf B_c\cdot(\mathbf{r\times P})
+U[V_l(r)+V_{so}(\mathbf{L}_r,\mathbf{S})]U^\dagger
-[V_l(r)+V_{so}(\mathbf{L}_r,\mathbf{S})]$
are small corrections which can be neglected
because of the following reasons:
In the parameter regime we are focusing on,
the diamagnetic contribution of the gradient field,
$\mathbf A_l(\mathbf{R+r})^2$, is small compared
to the one of the constant Ioffe field,
$\mathbf{A}_c(\mathbf{r})^2$.
The second contribution of $H_\text{corr}$ is negligible within our
adiabatic approach since $\langle \mathbf P/M\rangle$ becomes negligible
for ultracold temperatures compared to the relative motion
$\langle \mathbf p/m\rangle$. Finally, the remaining terms couple to
remote electronic states only and are therefore irrelevant.

The magnetic moments of the particles are
connected to the electronic spin $\mathbf{S}$ and the nuclear spin
$\mathbf{\Sigma}$ according to $\mbox{\boldmath$\mu$}_e=-\mathbf{S}$
and $\mbox{\boldmath$\mu$}_c=-\frac{g_N}{2M_c}\mathbf{\Sigma}$, with
$g_N$ being the nuclear $g$-factor; because of the large nuclear mass,
the term involving $\mbox{\boldmath$\mu$}_c$ is neglected
in the following.
We remark that the $Z$-component of the c.m.\ momentum commutes with the
Hamiltonian (\ref{eq:hamfinaluni}); hence the longitudinal motion can be
integrated out by employing plane waves $|K_Z\rangle=\exp(-iK_ZZ)$.
In order to solve the remaining coupled Schr\"odinger equation, we
employ a Born-Oppenheimer separation of the c.m.\ motion and the
electronic degrees of freedom by projecting Eq.~(\ref{eq:hamfinaluni})
on the electronic eigenfunctions $\psi_\kappa(\mathbf r;\mathbf R)$
that parametrically depend on the c.m.\ coordinates.
We are thereby led to a set of decoupled differential equations governing
the adiabatic c.m.\ motion within the individual two-dimensional energy
surfaces $E_\kappa(\mathbf R)$, i.e., the surfaces $E_\kappa(\mathbf R)$
serve as potentials for the c.m.\ motion of the atom.
The non-adiabatic (off-diagonal) coupling terms $\Delta T$
that arise within this procedure in the kinetic energy term
can be neglected in our parameter regime
since they are suppressed by the splitting between adjacent energy
surfaces \cite{hezel:223001}.
As will be shown in Sec.~\ref{sec:approximation},
the latter is proportional to the Ioffe field strength $B$,
i.e., the non-adiabatic couplings are proportional to powers of
$1/B$.

The electronic eigenfunctions and energies
are found by a standard basis set method utilizing the field-free
eigenfunctions $|\kappa\rangle=|njm_jls\rangle$ of $H_A$ whose
spin and angular parts $|jm_jls\rangle$ are given by the spin-orbit coupled
generalized spherical harmonics $\mathcal Y_{j,m_j,l}$ \cite{friedrich98}.
For the radial degree of freedom, a discrete variable representation (DVR) based on
generalized Laguerre polynomials is employed \cite{PhysRevA.67.042708}.
The latter provides a non-uniform grid for the radial coordinate
which is more dense close to the origin and hence
especially suited for representing radial Rydberg wave functions.
Since in the DVR scheme the potential matrix element evaluation
is equivalent to a Gaussian quadrature rule, representing
the Hamiltonian (\ref{eq:hamfinaluni})
-- especially $V_l(r)$ and the derivative terms arising from the
momentum operator $\mathbf p$ -- becomes particularly efficient.
The numerical diagonalization of the resulting Hamiltonian matrix
(in the limit $\mathbf P\rightarrow 0$) then yields the electronic
eigenfunctions $\psi_\kappa(\mathbf r;\mathbf R)$  and energies
$E_\kappa(\mathbf R)$ which
both parametrically depend on the c.m.\ coordinates $\mathbf R$.
Convergence is ensured by appropriately choosing the size of the
field-free basis as well as the underlying DVR grid size.

\begin{figure}
\includegraphics[width=8.5cm]{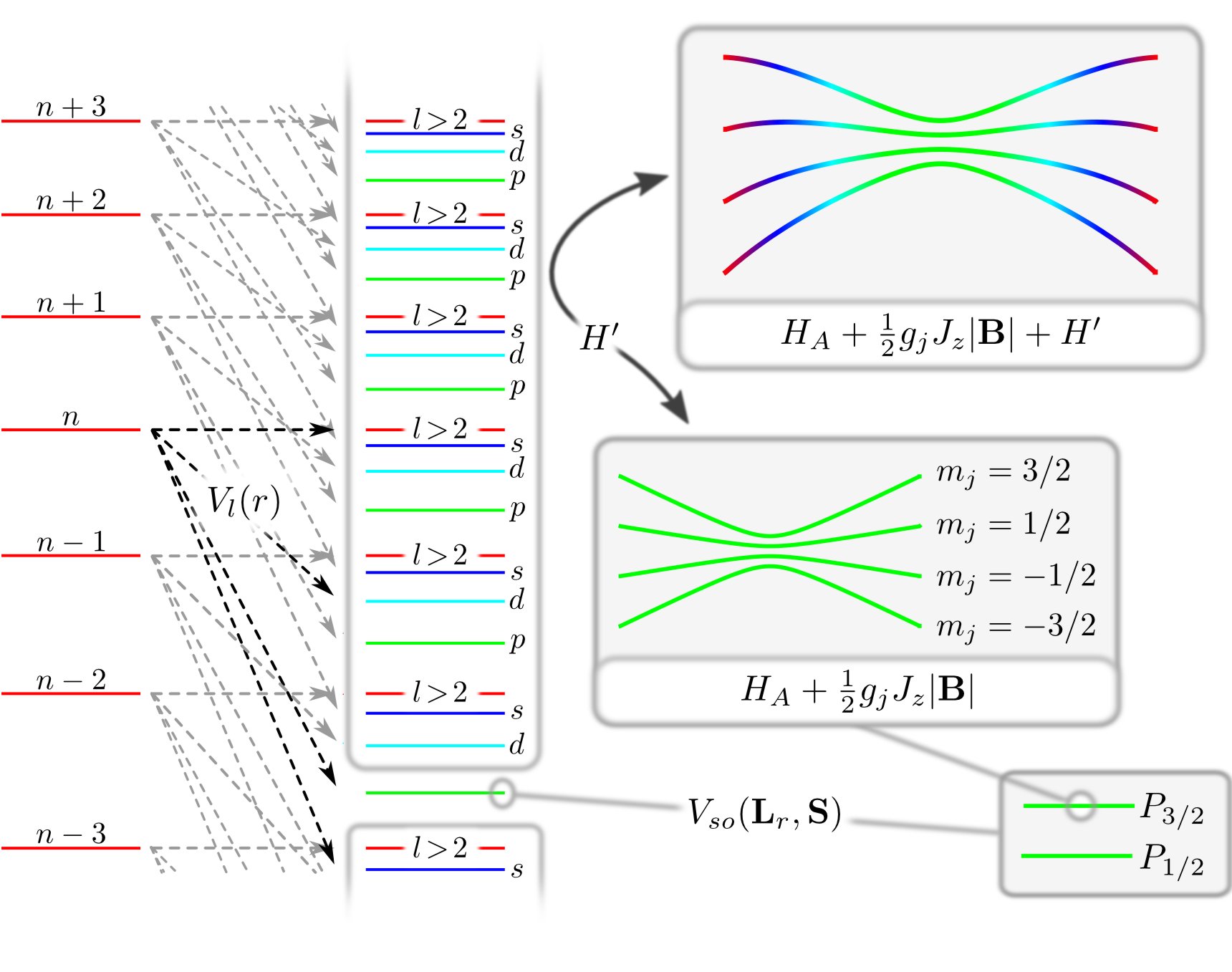}
\caption{(Color online) Schematic overview of the energy scales and
coupling terms involved.
Starting from the degenerate hydrogen energy spectrum on the left,
it is shown how the quantum defect -- modeled by the potential
$V_l(r)$ -- separates the low angular momentum states.
The spin-orbit coupling $V_{so}(\mathbf L_r,\mathbf S)$ then yields
the fine structure splitting for fixed $l$.
Within a given $j$-manifold, the Hamiltonian
$H_A+\frac{1}{2}g_jm_j|\mathbf B|$ resembles the coupling
of a point-like particle to the magnetic field $\mathbf B(\mathbf R)$.
The two-body character of the Rydberg atom, which is represented
by $H'$, only contributes if energetically remote
levels are considered as well: it admixes states of different
$n$, $l$, $j$, and $m_j$ thereby qualitatively changing the shape of the
surfaces.}
\label{fig:scheme}
\end{figure}

\section{Analytical Approach}
\label{sec:approximation}
While the above described numerical treatment of the electronic
Hamiltonian offers accurate results, we derive in this section
analytical but approximative expressions for the adiabatic energy
surfaces, which provides us with a profound understanding of the
underlying physics.
We start by considering only a single fine structure manifold,
i.e., fixed total angular momentum $j$ for given $l$.
Such an assumption is motivated by the fact that
the fine structure dominates over
the Zeeman splitting for the field strengths we are interested in,
cf.\ Fig.~\ref{fig:scheme}.
Within this regime, the contribution
$\mathbf{A}_l(\mathbf{R+r})\cdot\mathbf{p}=
G(XY+Xy+xY+xy)p_z$
of Hamiltonian (\ref{eq:hamfinaluni}) can be simplified as
follows.
We rewrite
\begin{equation}
\label{eq:lx}
 yp_z
= \frac{1}{2}(yp_z-zp_y)-\frac{i}{2}[yz,H_A]+
\frac{i}{2}[yz,V_l(r)+V_{so}(\mathbf{L}_r,\mathbf{S})]
\end{equation}
bearing in mind that only the action of any
involved operator within a single $j$-manifold
is considered.
The first commutator in the above equation then vanishes due
to the degeneracy of the eigenenergies of $H_A$ and since
no coupling to different $n$-, $l$-, or $j$-states is considered.
Likewise, the second commutator in Eq.~(\ref{eq:lx}) vanishes
since neither $V_l(r)$ nor $V_{so}(\mathbf{L}_r,\mathbf{S})$
depend on the magnetic quantum number $m_j$.
Hence, within a given $j$-manifold, the electronic part of
Hamiltonian (\ref{eq:hamfinaluni}) can be approximated by
\begin{equation}
\label{eq:helec}
H_e=H_A+\frac{1}{2}\left[\mathbf{L}_r+2\mathbf{S}\right]\cdot
\mathbf B(\mathbf R)+GXYp_z+H_r\,,
\end{equation}
where we substituted
$yp_z\rightarrow\frac{1}{2}L_x$ and similarly
$xp_z\rightarrow-\frac{1}{2}L_y$.
The contribution
$H_r=\mathbf{A}_l(\mathbf{r})\cdot\mathbf{p}+
\mathbf{B}_l(\mathbf{r})\cdot\mathbf{S}+
\tfrac{1}{2}\mathbf{A}_c(\mathbf{r})^2$
only depends on the relative coordinate and
-- as we will show later --
for a wide range of field strengths can approximately be
regarded as a mere energy offset to the electronic energy
surfaces; we will restrict ourselves to this regime
and hence omit $H_r$ in the following.

\begin{table}
\caption{Coefficients of the linear fit of $C_i(n)=C_i^{(0)}+C_i^{(1)}n$
in the range $35\le n\le45$ for the $nS$, $nP$, and $nD$ states of the
$^{87}$Rb atom.
Note that negative magnetic quantum numbers $m_j$ yield the same results
as their positive counterparts and are consequently omitted.
The fitted $C_i$ are calculated using Eq.~(\ref{eq:cgxy}) with
$n'\in[n-10,n+10]$.\label{tab:gxy}}
\begin{ruledtabular}
\begin{tabular}{c c c c c}
State&$C_x^{(0)}$&$C_x^{(1)}$&$C_z^{(0)}$&$C_z^{(1)}$\\
\hline
$S_{1/2},m_j=1/2$&-0.4813&-0.00027&-0.4813&-0.00027\\
$P_{1/2},m_j=1/2$&-0.4484&-0.00148&-0.4484&-0.00148\\
$P_{3/2},m_j=1/2$&-0.4541&-0.00152&-0.4316&-0.00164\\
$P_{3/2},m_j=3/2$&-0.4391&-0.00160&-0.4616&-0.00149\\
$D_{3/2},m_j=1/2$&-0.4570&-0.00069&-0.4326&-0.00011\\
$D_{3/2},m_j=3/2$&-0.4407&-0.00030&-0.4652&-0.00088\\
$D_{5/2},m_j=1/2$&-0.4570&-0.00073&-0.4287&-0.00006\\
$D_{5/2},m_j=3/2$&-0.4500&-0.00057&-0.4429&-0.00040\\
$D_{5/2},m_j=5/2$&-0.4358&-0.00023&-0.4712&-0.00107
\end{tabular}
\end{ruledtabular}
\end{table}

The first two terms of Hamiltonian~(\ref{eq:helec}) can be
diagonalized analytically by applying the spatially dependent
transformation
\begin{equation}
 U_r=e^{-i \gamma (L_x+S_x)}e^{-i \beta (L_y+S_y)}
\label{eq:Urot}
\end{equation}
that rotates the $z$-axis into the local magnetic field
direction; $\gamma$ and $\beta$ denote the rotation angles:
\begin{align}
\sin\gamma& =\frac{-GY}{\sqrt{B^2+G^2(X^2+Y^2)}}\,,\\
\sin\beta& =\frac{-GX}{\sqrt{B^2+G^2X^2}}\,,\\
\cos\gamma& =\frac{\sqrt{B^2+G^2X^2}}{\sqrt{B^2+G^2(X^2+Y^2)}}\,,\\
\cos\beta& =\frac{B}{\sqrt{B^2+G^2X^2}}\,.
\end{align}
The transformed Hamiltonian becomes
\begin{equation}
 U_r H_e U_r^\dagger =H_A+\tfrac{1}{2}g_jJ_z\sqrt{B^2+G^2(X^2+Y^2)}\,,
+H'
\label{eq:htrans}
\end{equation}
with $g_j=\frac{3}{2}+\frac{s(s+1)-l(l+1)}{2j(j+1)}$,
$U_r H_A U_r^\dagger=H_A$, and
$H'=GXYU_r p_z  U_r^\dagger$.
Like for ground state atoms, the second term of Eq.~(\ref{eq:htrans})
represents the coupling of a point-like particle to the magnetic field
via its magnetic moment $\boldsymbol{\mu}=\frac{1}{2}\mathbf L_r
+\mathbf S$.

As depicted in Fig.~\ref{fig:scheme}, $H'$ couples to
different $n$, $l$, $j$, and $m_j$ and hence vanishes within one
$j$-manifold.
The first two terms of Eq.~(\ref{eq:htrans}), on the
other hand,
are diagonal, giving rise to the electronic potential energy surface 
\begin{equation}\label{eq:ekappa0}
E_\kappa^{(0)}(\mathbf R) =E_\kappa^{el}+\tfrac{1}{2}g_jm_j
\sqrt{B^2+G^2(X^2+Y^2)}
\end{equation}
for a given electronic state $|\kappa\rangle=|njm_jls\rangle$.
Note that such a state refers to the \emph{rotated} frame of reference.
Only there, $m_j$ constitutes a good quantum number; in the laboratory
frame of reference $m_j$ is not conserved. 
The surfaces Eq.~(\ref{eq:ekappa0}) are rotationally symmetric around the $Z$-axis and confining
for $m_j>0$. For small radii ($\rho=\sqrt{X^2+Y^2}\ll B/G$) an expansion
up to second order yields a harmonic potential
\begin{equation}
E_\kappa^{(0)}(\rho)\approx E_\kappa^{el}+\tfrac{1}{2}g_jm_jB+
\frac{1}{2}M\omega^2\rho^2
\label{eq:harmonic}
\end{equation}
with the trap frequency defined by $\omega=G\sqrt{\frac{g_jm_j}{2MB}}$
while we find a linear behavior
$E_\kappa^{(0)}(\rho)\approx E_\kappa^{el}+\tfrac{1}{2}g_jm_jG\rho$
when the center of mass is far from the $Z$-axis ($\rho\gg B/G$).
In the harmonic part of the potential, the c.m.\ energies are thus given by
\begin{equation}
 E_{\kappa,\nu}^{cm}=E_\kappa^{(0)}(0)+(\nu+1)\omega\:,\;\nu=
\nu_x+\nu_y\in\mathbb{N}
\end{equation}
with a splitting of $\omega$ between adjacent c.m.\ states;
see Sec.~\ref{sec:cmwave} for a more detailed discussion.
The separation between adjacent electronic energy surfaces
at the origin, on the other hand, is given by
$\Delta E_\kappa=\tfrac{1}{2}g_jB$.
The size of the c.m.\ ground state ($\nu=0$) in such a harmonic potential
evaluates to
$\langle \rho \rangle=\sqrt{\pi}/2\sqrt{M\omega}$
\cite{hezel:053417}.

\begin{figure}
\includegraphics[width=8.5cm]{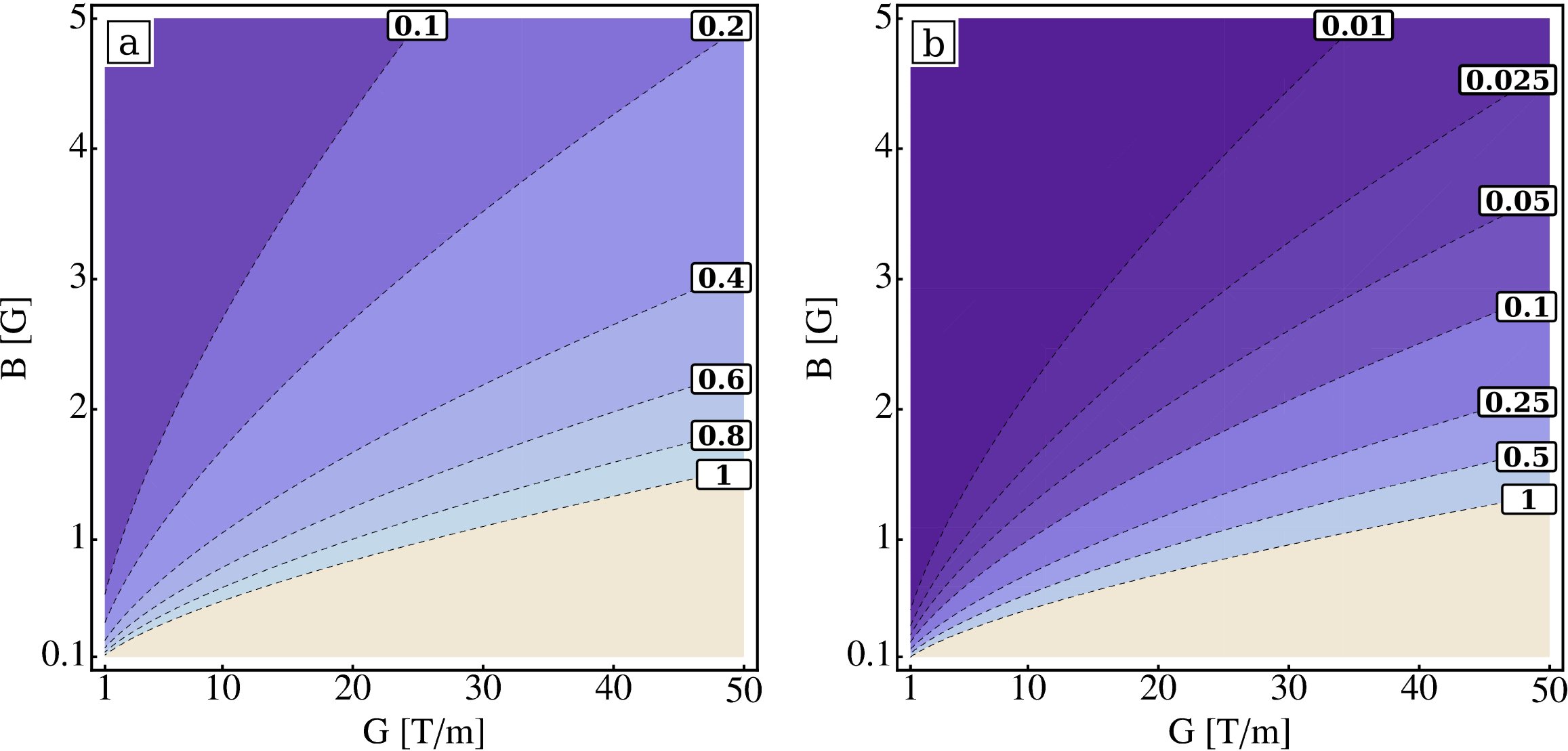}
\caption{(Color online) Range of validity for deriving Eq.~(\ref{eq:Xpm}).
(a) $2G^2X^2/B^2\ll 1$ resulting in
$X\ll B/\sqrt{2}G$. In the figure, the value
$X_+/(B/\sqrt{2}G)$ is shown which should be $\ll 1$.
Figure (b) shows $G^2X_0^2/B^2$ which should be $\ll 1$
as well. In both cases $l=2, j=m_j=5/2$ is used.}
\label{fig:valid}
\end{figure}

The remaining term $H'$ of Hamiltonian (\ref{eq:htrans}) 
can be treated perturbatively.
While it vanishes in first order, second-order perturbation theory
yields
\begin{align}
 E_\kappa^{(2)}(\mathbf R)&=G^2X^2Y^2\sum_{\kappa'\neq\kappa}
\frac{\big|\langle \kappa|U_rp_zU_r^\dagger|\kappa'\rangle\big|^2}
{E_\kappa^{el}-E_{\kappa'}^{el}}
\label{eq:gxy_sum}\\
&\approx G^2X^2Y^2\sum_{\kappa'\neq\kappa}
(E_\kappa^{el}-E_{\kappa'}^{el})\times\big|\langle \kappa|
U_rzU_r^\dagger|\kappa'\rangle\big|^2
\label{eq:gxy}
\end{align}
where $\kappa=|njm_jls\rangle$ are the eigenstates of the transformed
Hamiltonian $U_r H_e U_r^\dagger-H'$, cf.\ Eq.~(\ref{eq:htrans}).
Since $E_\kappa^{(0)}(\mathbf R)$ resembles the confinement of a
ground state atom, we attribute
$E_\kappa^{(2)}(\mathbf R)$ to the composite nature of the Rydberg
atom, i.e., the fact that it consists of a Rydberg electron far
apart from its ionic core. Like the magnetic field itself,
$E_\kappa^{(2)}(\mathbf R)$ shows no continuous azimuthal symmetry
but rather a discrete one. 

Equation (\ref{eq:gxy}) is derived by employing
$U_rp_zU_r^\dagger=p_x\sin\beta-p_y\sin\gamma\cos\beta+
p_z\cos\gamma\cos\beta$
and $\mathbf p=i[H_A,\mathbf r]-i[V_l(r)+V_{so},\mathbf r]\approx
i[H_A,\mathbf r]$
\footnote{The terms involving $[V_l(r)+V_{so},\mathbf r]$
in Eq.~(\ref{eq:gxy_sum}) give rise to
$|\langle\kappa|r(V_l(r)-V_{l'}(r))|\kappa'\rangle|^2/
(E_\kappa^{el}-E_{\kappa'}^{el})$ and
$|\langle\kappa|r(V_{so}-V_{so}')|\kappa'\rangle|^2/
(E_\kappa^{el}-E_{\kappa'}^{el})$
which both can be neglected compared to the leading contribution of
$(E_\kappa^{el}-E_{\kappa'}^{el})\times|\langle\kappa|r|\kappa'\rangle|^2$.}.
Expanding the modulus square in Eq.~(\ref{eq:gxy}),
one obtains mixed terms of the form
$\langle\kappa|x|\kappa'\rangle^*\langle\kappa|y|\kappa'\rangle+
\langle\kappa|x|\kappa'\rangle\langle\kappa|y|\kappa'\rangle^*$.
Employing the standard basis of spherical harmonics and
consequently using $\langle\kappa|x|\kappa'\rangle\in \mathbb R$ as well as
$\langle\kappa|y|\kappa'\rangle=-\langle\kappa|y|\kappa'\rangle^*$,
this sum vanishes.
The matrix element of $z$ obeys
a different selection rule, namely, $\Delta m_l=0$ opposed to $\Delta m_l=\pm1$
of $x$ and $y$; hence, mixed terms involving $\langle\kappa|z|\kappa'\rangle$
vanish as well.
Consequently, only the matrix elements
$|\langle\kappa|x|\kappa'\rangle|^2$, $|\langle\kappa|y|\kappa'\rangle|^2$,
and $|\langle\kappa|z|\kappa'\rangle|^2$ remain and
the second order energy contribution can be parametrized as
\begin{align}
 E_\kappa^{(2)}(\mathbf R)={}&G^2X^2Y^2\big(C_x\sin^2\!\beta
\nonumber\\
&+C_y\sin^2\!\gamma\cos^2\!\beta+C_z\cos^2\!\gamma\cos^2\!\beta\big)
\label{eq:gxy_c}\\
={}&C_z G^2X^2Y^2\nonumber\\
&\times\Big[1+\frac{C_x-C_z}{C_z}\left( \sin^2\!\beta+\sin^2\!\gamma
\cos^2\!\beta\right)\Big].
\label{eq:gxydiff}
\end{align}
with
\begin{equation}
C_i=\sum_{\kappa'\ne \kappa}(E_{\kappa}^{el}-E_{\kappa'}^{el})|
\langle\kappa|x_i|\kappa'\rangle|^2
\label{eq:cgxy}
\end{equation}
where $C_x=C_y=C_z\equiv C$ for $l=0$ and $C_x=C_y$ otherwise
(since $|\langle\kappa|x|\kappa'\rangle|=|\langle\kappa|y|\kappa'\rangle|$).
Note that the parameters $C_i$ depend on the state $\kappa$ under
investigation.
Since $E_\kappa^{el}-E_{\kappa'}^{el}\propto n^{-3}$ and
$|\langle \kappa|x_i|\kappa'\rangle|^2\propto n^4$, a linear
scaling of $E_\kappa^{(2)}(\mathbf R)$ with the quantum number $n$ is
anticipated, i.e., $C_i(n)=C_i^{(0)}+C_i^{(1)}n$.
Resulting from a fit of calculated $C_i$ values within the range
$35\le n\le45$, in Tab.~\ref{tab:gxy} the coefficients $C_i^{(j)}$ are
tabulated for the $nS$, $nP$, and $nD$ states of the $^{87}$Rb atom.
All considered states show a similar behavior: The magnitude of $C_i$ is
close to $-1/2$ and shows a rather weak $n$-dependence.
In particular, $C_x\approx C_z$ and therefore
$E_\kappa^{(2)}(\mathbf R)\approx C_z\cdot G^2X^2Y^2$,
cf.\ Eq.~(\ref{eq:gxydiff}).
We remark that for smaller $n$, $C_i(n)$ deviates from the linear behavior
in favor of a more rapid decrease.

\begin{figure*}
\includegraphics[width=17.75cm]{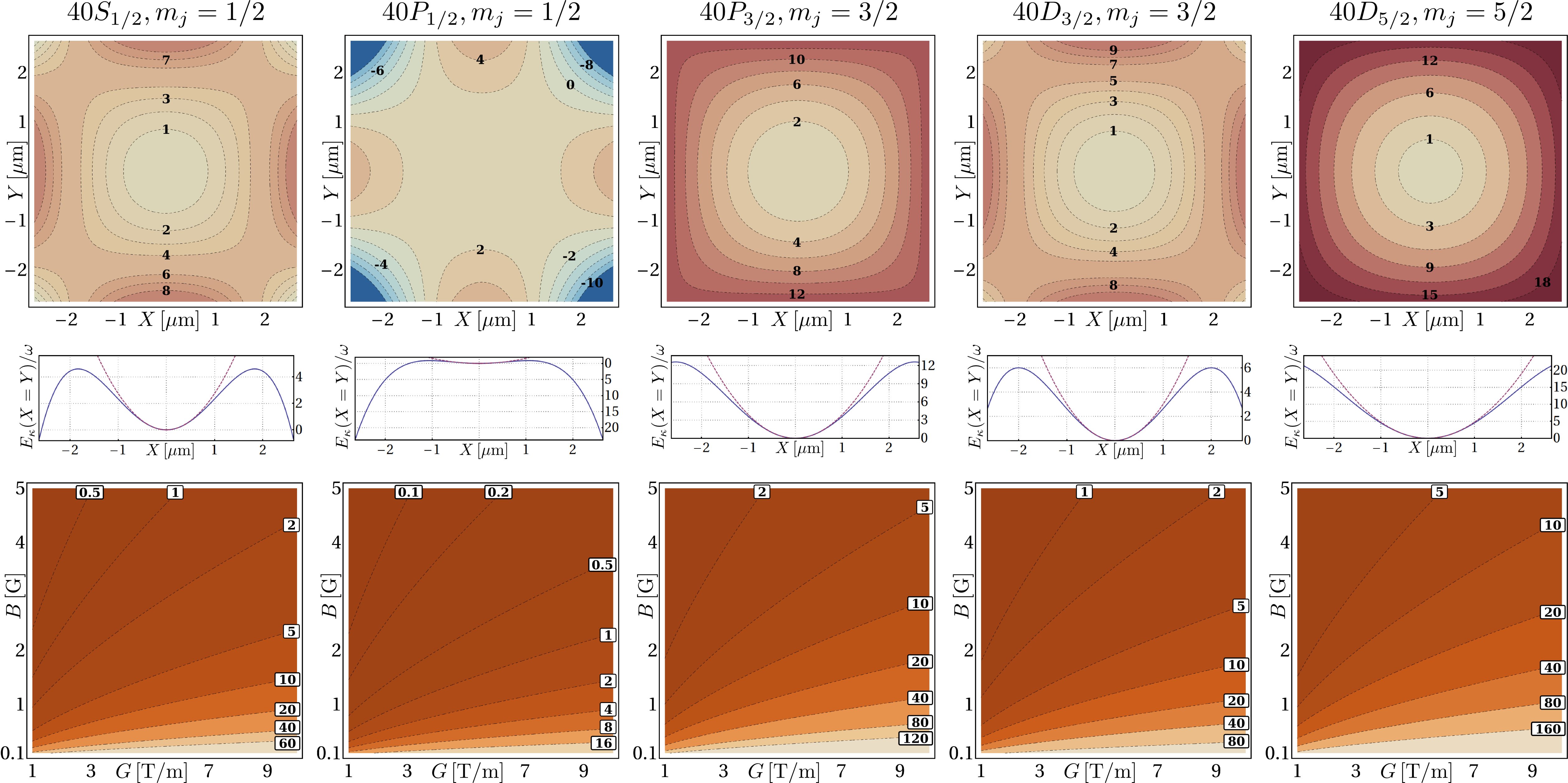}
\caption{(Color online) First Row: Contour plots of the electronic potential energy surfaces
$E_\kappa(\mathbf R)$ of the $40S_{1/2}$ (first column),
$40P_{1/2}$ (second column), $40P_{3/2}$ (third column),
$40D_{3/2}$ (fourth column), and $40D_{5/2}$ (fifth column)
 states with $m_j=j$ for the magnetic field configuration $B=1$ G, $G=2.5$ T/m.
Second Row: Section along $X=Y$ of the same energy surfaces;
the contribution of $E_\kappa^{(0)}(\mathbf R)$ (dashed lines) is shown in
addition.
Third Row: Depth of the potentials as given by Eq.~(\ref{eq:de_exact})
as a function of the magnetic field configuration.
For all subfigures, the energy scale is given in terms of the
trap frequency $\omega=G\sqrt{g_jm_j/2MB}$.}
\label{fig:surfaces}
\end{figure*}

In the last part of this section, let us reconsider
the adiabatic energy surfaces for the c.m.\ motion,
including now the contribution of $H'$.
That is, we investigate the approximate, but analytical solutions
\begin{equation}
 E_\kappa(\mathbf R)\equiv E_\kappa^{(0)}(\mathbf R)+
 E_\kappa^{(2)}(\mathbf R)-E_\kappa^{el},
\label{eq:ealpha}
\end{equation}
of Hamiltonian (\ref{eq:helec}).
In particular, we concentrate on the diagonal of the surfaces ($X=Y$)
where $E_\kappa^{(2)}(\mathbf R)$ is maximal.
The approximation $C_x=C_z$ (which is exact for $nS_{1/2}$ states)
then yields
\begin{equation}
 E_\kappa(X=Y)=\frac{1}{2}g_jm_j\sqrt{B^2+2G^2X^2}+C_z G^2X^4\,,
\end{equation}
which shows only a \emph{local} minimum at the origin
since the surface drops off for large c.m.\ coordinates
when $E_\kappa^{(2)}(\mathbf R)$ dominates (note that $C_z<0$),
see also Fig.~\ref{fig:surfaces}.
The positions of the maxima which enclose the minimum are
approximately given by
\begin{equation}
 X_\pm\approx\pm\left(\frac{G^2}{B^2}-\frac{4BC_z}{g_jm_j}\right)^{-1/2}
\approx\pm X_0\left(1-\frac{G^2}{2B^2}X_0^2\right)
\label{eq:Xpm}
\end{equation}
with the length scale $X_0=\sqrt{\frac{g_jm_j}{4B|C_z|}}$ only depending on
the Ioffe field strength.
The depth of the potential well associated with
the minimum correspondingly evaluates
to
\begin{align}
 \Delta E_\kappa={}&E_\kappa(X=Y=X_+)-E_\kappa(0)\label{eq:de_exact}\\
\approx{}&
\frac{1}{2}g_jm_j\frac{G^2X_0^2}{B}
\left(1-\frac{G^2X_0^2}{B^2}\right)
\nonumber\\
+&C_z G^2X_0^4\left(1-2\frac{G^2X_0^2}{B^2}\right).
\end{align}
Note that the first approximation in Eq.~(\ref{eq:Xpm}) holds
for $2G^2X^2/B^2\ll 1$ and the second one for
$G^2X_0^2/B^2\ll 1$. The corresponding range of validity is illustrated in
Fig.~\ref{fig:valid}: For a Ioffe field strength of $B=1$ G, the
above approximations hold for gradients up to $10$ T/m; at higher $B$
even larger gradients are eligible.

\section{Trapping potentials}
\label{sec:surfaces}
In the following section we are going to discuss the calculated 
electronic potential energy surfaces for the $nS$, $nP$, and $nD$ 
states of the $^{87}$Rb atom in detail. In particular, the range 
of validity of the above derived analytic expression 
[Eq.~(\ref{eq:ealpha})] is demonstrated. As a general example, we 
address the magnetic field configuration $B=1$\,G, $G=2.5\,$Tm$^{-1}$, 
which yields a trap frequency of $\omega=2\pi\times319$ Hz. A 
similar field configuration is also found in current experiments 
\cite{Loew2007}. 

\begin{figure}
\includegraphics{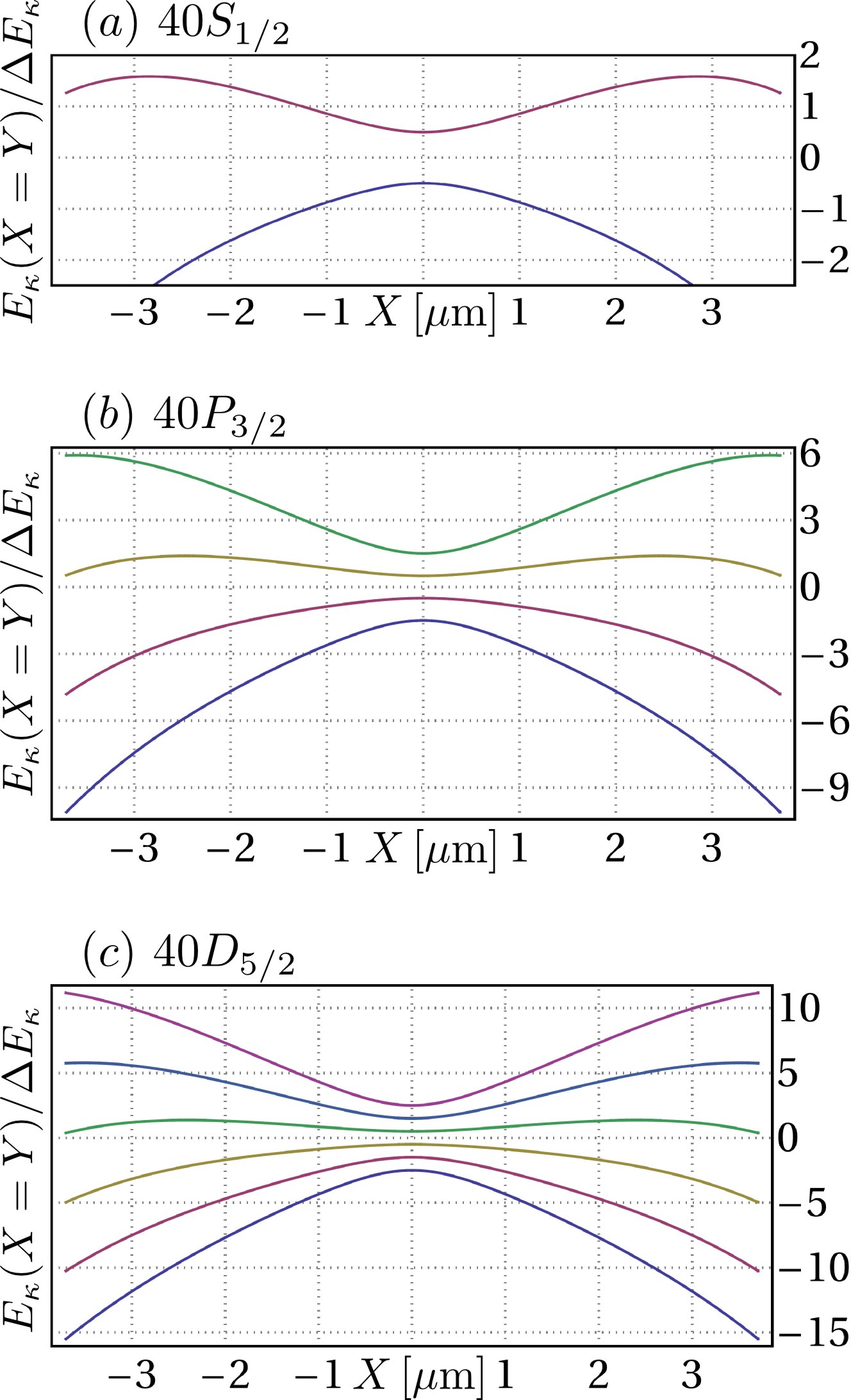}
\caption{(Color online) Sections along $X=Y$ of the energy surfaces of the
(a) $40S_{1/2}$, (b) $40P_{3/2}$, and (c) $40D_{5/2}$ states
for the field configuration $B=0.1$ G, $G=10$ T/m,
which yields a trap frequency of $\omega/\sqrt{g_jm_j}	=2\pi\times4$ kHz.}
\label{fig:spec}
\end{figure}

In Figure \ref{fig:surfaces} the electronic potential energy 
surfaces $E_\kappa(\mathbf R)$ of the 
$40S_{1/2}$, $40P_{1/2}$, $40P_{3/2}$, $40D_{3/2}$, 
and $40D_{5/2}$ states with $m_j=j$ are illustrated. In addition,
also sections along $X=Y$ of these surfaces are provided. On a first 
glance, the energy surfaces originating from different electronic 
states seem to differ quite substantially. However, qualitatively 
they are very similar, as we are going to argue in the following. 
For all surfaces presented in Fig.~\ref{fig:surfaces}, the contribution 
of the composite character of the Rydberg atom, i.e., 
$E_\kappa^{(2)}(\mathbf R)$ changes the azimuthal symmetry of 
$E_\kappa^{(0)}(\mathbf R)$ into a four-fold one. Moreover, the 
interplay between the harmonic confinement $E_\kappa^{(0)}(\mathbf R)$ 
and the unbounded contribution $E_\kappa^{(2)}(\mathbf R)$ 
gives rise to a finite trap depth along the diagonal $X=Y$; see 
second row of Fig.~\ref{fig:surfaces}. Since the coefficient $C_z$ 
of $E_\kappa^{(2)}(\mathbf R)$ is approximately of the same magnitude 
for all states considered, cf.\ Tab.~\ref{tab:gxy}, the trap depth 
depends on the magnitude of the magnetic moment 
$\boldsymbol{\mu}\propto g_j\mathbf J$. Consequently, the $j=m_j=l+1/2$ 
electronic states show a deeper confinement than their $j=m_j=l-1/2$ 
counterparts and the depth increases further with increasing orbital 
angular momentum $l$. For the examples given in Fig.~\ref{fig:surfaces} 
this means that the quadratic approximation to the trapping potential 
for the $40S_{1/2}$ state is already violated at about two oscillator 
energies, while for the $40D_{5/2}$ it is fine up to $10\,\omega$. This 
trend is confirmed in the third row of Fig.~\ref{fig:surfaces} where 
the depth of the potential as a function of the field configuration is 
displayed: For the $40D_{5/2}$ state, the trap depth easily exceeds 
$100\, \omega$ within the given parameter range, while in the case of 
the $40S_{1/2}$ state there is a substantial regime of field strengths 
where not a single center of mass state can be confined, i.e., the trap 
depth being $<1\, \omega$. Nevertheless, also for the $40S_{1/2}$ Rydberg 
state the field parameters $B$ and $G$ can be adjusted such that trapping 
is possible, i.e., the trap depth being much larger than the trap frequency. 
Similarly, the trapping potential of the $40D_{5/2}$ Rydberg state can be 
chosen very shallow by going to sufficiently strong Ioffe fields. We remark 
that the results presented in Fig.~\ref{fig:surfaces} are given in units of 
the trap frequency $\omega=G\sqrt{g_jm_j/2MB}$; the latter holds, of course, 
only near the origin. For larger radii, the contribution 
$E_\kappa^{(2)}(\mathbf R)$ flattens the potential resulting in smaller trap 
frequencies and hence in a higher number of center of mass states that can be 
confined. Moreover, for very high gradients the harmonic expansion of the 
magnetic field strength becomes progressively worse.

As can be deduced from the third row of Fig.~\ref{fig:surfaces}, 
increasing the relative strength of the field gradient, i.e., either 
increasing $G$ directly or decreasing the offset field $B$ for fixed 
$G$, leads to a larger number of bound center of mass states -- 
independently of the state under consideration. However, since the 
anti-trapping contribution $E_\kappa^{(2)}(\mathbf R)$ quadratically 
increases with the field gradient $G$, we expect this trend to reverse 
for sufficiently high gradients. Indeed, for a Ioffe field of $B=1\,$G 
the trap depth starts to decrease for field gradients 
$G\gtrsim 200\,\text{Tm}^{-1}$; for $B=0.1\,$G this trend already 
starts at $G\gtrsim 5\,\text{Tm}^{-1}$. Similarly, for a fixed field 
gradient together with a decreasing offset field $B$ we find a decrease 
of the trap depth if $B\lesssim 0.15\,$G or $B\lesssim 0.03\,$G for 
$G=10\,\text{Tm}^{-1}$ and $G=1\,\text{Tm}^{-1}$, respectively.

In the following, let us investigate the question if electronic energy
surfaces belonging to different states intersect each other.
In Refs.~\cite{hezel:053417,hezel:223001}, where high angular momentum
states are considered, this issue is essential: there,
the high level of degeneracy of
the system leads to non-adiabatic crossings of the surfaces. As a
consequence, only the circular electronic state ($m_l=l=n-1$) provides
stable trapping.
For the low angular momentum states of $^{87}$Rb we are considering here, however,
the fine structure splitting for different $j$ and the varying quantum defect
for different $l$ separate the energy surfaces by lifting the degeneracy,
therefore preventing their crossing.
As a result, it is sufficient in our case to investigate the energy surface
spectrum for fixed $j$, i.e., only as a function of the magnetic
quantum number.
In Figure \ref{fig:spec}, sections along the diagonal of the energy surfaces
of the multiplets of the $40S_{1/2}$, $40P_{3/2}$, and $40D_{5/2}$ states are
presented.
In order to show a strong spatial dependence, we choose an extreme
case concerning the ratio of the Ioffe field compared to the gradient field,
namely, $B=0.1$ G, $G=10$ T/m.
Even for such a high gradient (of course much higher gradients can be achieved
on atom chips),
the energy surfaces remain well separated
with a minimum distance of $\frac{1}{2}g_j B$ at the trap center.
Hence, each surface can be considered separately for trapping
and our adiabatic approach is not limited by non-adiabatic interactions.
Note that the anti-trapping of $m_j<0$ states
is even enhanced by the contribution $E_\kappa^{(2)}(\mathbf R)$.

\begin{figure}
\includegraphics[width=7cm]{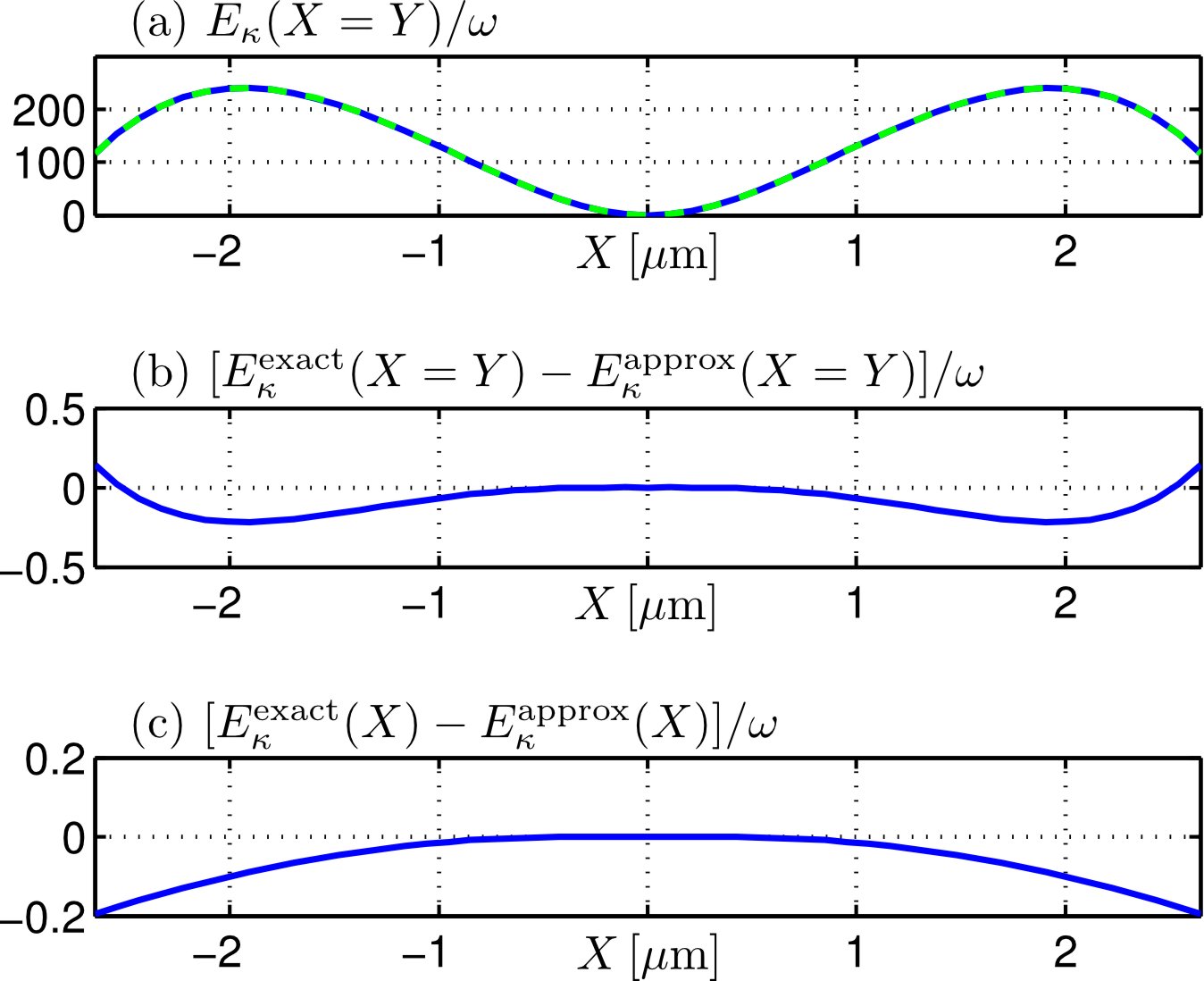}
\caption{(Color online) (a) Section along $X=Y$ of the energy surface of the
$40D_{5/2}, m_j=5/2$ state
for the field configuration $B=1$ G, $G=100$ T/m
which yields a trap frequency of $\omega=2\pi\times22.1$ kHz near
the origin. The solid line is yield by the numerical diagonalization of
Hamiltonian (\ref{eq:hamfinaluni}), the dashed one by Eq.~(\ref{eq:ealpha}).
The surfaces have been offset to zero at the origin.
(b) Difference (in terms of the
trap frequency $\omega$) between the analytic expression Eq.~(\ref{eq:ealpha})
and the result of the 
numerical diagonalization of Hamiltonian (\ref{eq:hamfinaluni}).
(c) Same as (b) but for $Y=0$ rather than $X=Y$.}
\label{fig:diff}
\end{figure}

The above investigations employ the analytical expression
Eq.~(\ref{eq:ealpha}) rather than the exact numerical solutions of
Hamiltonian (\ref{eq:hamfinaluni}).
Hence an estimation of the range of validity of our results is
necessary.
To this end, we provide in Fig.~\ref{fig:diff}(a) a comparison between
the analytical expression according to Eq.~(\ref{eq:ealpha}) and
the numerical diagonalization of Hamiltonian (\ref{eq:hamfinaluni})
for the `extreme' field configuration of $B=1$ G, $G=100$ T/m
(the quantitative agreement improves for smaller gradients /
larger Ioffe fields).
As one can observe, even for such a strong gradient Eq.~(\ref{eq:ealpha})
yields satisfactory results: The deviation within the spatial range
considered is less than $0.2\,\omega$, cf.\ Fig.~\ref{fig:diff}(b),
$\omega$ being the splitting of
the c.m.\ states and representing the smallest energy scale of the system.
We remark that Fig.~\ref{fig:diff}(b) shows results along the diagonals $X=Y$
of the surfaces where the deviation is at maximum.
However, even along the axes,
where $E_\kappa^{(2)}(\mathbf R)$ vanishes, a perfect agreement cannot be
found, cf.\ Fig.~\ref{fig:diff}(c).
This residual deviation is due to the purely electronic terms $H_r$
of Hamiltonian (\ref{eq:hamfinaluni}) which have been neglected in
deriving Eq.~(\ref{eq:ealpha}). Although $H_r$ does not explicitly
depend on the c.m.\ coordinate $\mathbf R$, it introduces an implicit
c.m.~dependency by changing the electronic state. This can be easily
understood in the rotated frame of reference, i.e., after applying the
unitary transformation $U_r$: There one has to consider
$U_r H_r U_r^\dagger$ which introduces a c.m.~dependency explicitly.
We stress that for gradients weaker and/or Ioffe fields stronger than
in Fig.~\ref{fig:diff}, $U_r$ is closer to unity and therefore
the contribution of $H_r$ becomes even less important in these cases and
hence can be neglected.

\section{c.m.\ wave functions}
\label{sec:cmwave}
In this section we discuss the eigenfunctions
$\chi(\mathbf R)$ of the c.m.\ Hamiltonian
\begin{equation}
 H_{cm}=\frac{\mathbf P^2}{2 M}+E_\kappa(\mathbf R)
\label{eq:cmhamiltonian}
\end{equation}
where $E_\kappa(\mathbf R)$ are the previously calculated potentials.
In particular, we are going to elucidate the differences to the harmonic 
oscillator eigenstates, that are yielded by considering solely the `unperturbed'
potential $E_\kappa^{(0)}(\mathbf R)$ (throughout this section, the harmonic 
approximation Eq.~(\ref{eq:harmonic}) for the potential 
$E_\kappa^{(0)}(\mathbf R)$ is assumed).
The energies and eigenstates are then computed using second
order perturbation theory.
We remark that the validity of the harmonic approximation together with
the use of perturbation theory has been ensured by comparing with
the results obtained by the numerical diagonalization of
Hamiltonian (\ref{eq:cmhamiltonian}).

Before presenting our results, let us comment on the issue of the finite
radiative lifetime of Rydberg atoms which might spoil the experimental
observation of the c.m.\ motion.
The lifetime can be
parametrized as $\tau=\tau'(n-\delta_\kappa)^\gamma$
where one finds
$\tau'=1.43$ ns and $\gamma=2.94$ for $l=0$,
$\tau'=2.76$ ns and $\gamma=3.02$ for $l=1$, and
$\tau'=2.09$ ns and $\gamma=2.85$ for $l=2$ \cite{PhysRevA.65.031401}.
For the $40S_{1/2}$ Rydberg state, this yields a radiative
lifetime of $\tau=58\,\mu$s.
If we compare this to the typical time scale $\tau_\omega=2\pi/\omega$
of the c.m.\ motion, one finds that for the envisaged field
configuration ($B=1$ G, $G=2.5$ T/m) $\tau_\omega=3$ ms
is orders of magnitudes larger than the radiative lifetime,
which renders the resolution of the c.m.\ motion experimentally
impossible.
This drawback can be alleviated by several means. First of all,
one can consider higher principal quantum numbers $n$
which increases the lifetime
substantially. For example, the $60S_{1/2}$ state already
possesses a radiative lifetime of $\tau=206\,\mu$s.
The changes in the trapping potential, on the other hand,
are marginal as can be seen from the weak $n$-dependence
of the $C_z$ coefficient, cf.\ Tab.~\ref{tab:gxy};
note that $E_\kappa^{(0)}(\mathbf R)$ is $n$-independent.
Additionally to increasing $n$, one can augment the trap
frequency by increasing the gradient field and/or decreasing
the Ioffe field (which might necessitate atom chip traps
\cite{folman,fortagh:235}).
As an example, the field configuration
$B=0.1$ G and $G=50$ T/m yields $\tau_\omega=50\,\mu$s.
Furthermore, one might also employ the $nP_{3/2}$ Rydberg states
which possess a longer lifetime
($\tau=155\,\mu$s and $\tau=0.5$ ms for $n=40$ and $n=60$,
respectively) and at the same time cause a higher trap
frequency
($\tau_\omega=2$ ms and $\tau_\omega=35\,\mu$s for
$B=1$ G, $G=2.5$ T/m and $B=0.1$ G, $G=50$ T/m,
respectively).

As an illustrative example, let us investigate again
the $40S_{1/2}$ Rydberg state combined with the magnetic field parameters
$B=1$ G and $G=2.5$ T/m in the following,
despite the above mentioned restrictions.
In this case, the resulting trapping potential $E_\kappa(\mathbf R)$
confines only a very limited number of c.m.\ states, namely, twelve.
Consequently, already low c.m.\ excitations show an
appreciable deviation from the harmonic behavior
which makes the influence of the perturbative
effects of $E_\kappa^{(2)}(\mathbf R)$ particularly
visible.
For the case of $E_\kappa^{(0)}(\mathbf R)$ and small c.m.\ radii
one yields a harmonic potential.
In this case, the Hamiltonian (\ref{eq:cmhamiltonian}) decouples
in $X$ and $Y$, i.e., the total c.m.\ wave function can be written
as a product of two independent harmonic oscillator states
in $X$ and $Y$:
$\chi(\mathbf R)\equiv\chi_{\nu_x\nu_y}(\mathbf R)=\chi_{\nu_x}(X)
\cdot \chi_{\nu_y}(Y)$.
As a consequence, the corresponding energies only depend
on the sum of the individual c.m.\ excitations $\nu=\nu_x+\nu_y$
and show a $(\nu+1)$-fold degeneracy.

In anticipation of considering $E_\kappa^{(2)}(\mathbf R)$ as well,
it is advisable to employ adapted eigenstates
which account for the ${\bm C}_{4v}$
symmetry of Hamiltonian (\ref{eq:cmhamiltonian}) including
$E_\kappa^{(2)}(\mathbf R)$.
The decomposition of such symmetry adapted eigenstates 
in terms of the product states
$\chi_{\nu_x\nu_y}(\mathbf R)$
can be found in the fourth column of Tab.~\ref{tab:cmstates}
together with their corresponding symmetry label given in the second column.
Note that these states are still degenerate in case of the
potential $E_\kappa^{(0)}(\mathbf R)$, cf.\ third column of
Tab.~\ref{tab:cmstates}.
The inclusion of $E_\kappa^{(2)}(\mathbf R)$ lifts this degeneracy
by mixing states of equal symmetry according to the vanishing
integral rule \cite{bunkerjensen}: since $E_\kappa^{(2)}(\mathbf R)$
is of $A_1$ symmetry, i.e., being totally symmetric
the c.m.\ matrix element
$\langle\chi'|E_\kappa^{(2)}|\chi\rangle$
is only non-vanishing if $|\chi\rangle$ and $|\chi'\rangle$
possess the same symmetry.
Moreover, the perturbation of the form $\sim X^2Y^2$ yields
the selection rules $\Delta \nu_{x}\in\{0,\pm2\}$ and
$\Delta \nu_{y}\in\{0,\pm2\}$.
Inspecting the wave functions as obtained by
diagonalizing Hamiltonian (\ref{eq:cmhamiltonian}) within
a formerly degenerate $\nu$-manifold, both the symmetry constraints
as well as the selection rules become apparent;
see sixth column of Tab.~\ref{tab:cmstates}.

\begin{table*}
\caption{Third Column:
Energies of the c.m.\ Hamiltonian (\ref{eq:cmhamiltonian})
for the harmonic potential $E_\kappa^{(0)}(\mathbf R)$.
Fourth Column: Corresponding symmetry adapted eigenstates.
Fifth Column: Second order perturbation theory energies
including the contribution $E_\kappa^{(2)}(\mathbf R)$.
Sixth Column: Corresponding zero order perturbation theory
wave functions.
The symmetries of the states encountered are listed in the second column.
Seventh column: Tunneling probability $P_t$
for the energies given in the fifth column.
\label{tab:cmstates}}
\begin{ruledtabular}
\begin{tabular}{c c c c c c c}
\multirow{2}{*}{state no.}&\multirow{2}{*}{symmetry}&h.o.&symmetry adapted&perturbed&zero order&tunneling\\
&&energies& eigenstates & energies&eigenstates & probability\\
\hline
ground&$A_1$&$\omega$&$\chi_{00}$&0.9859\,$\omega$&$\chi_{00}$&$3.19\times 10^{-6}$\\
1&\multirow{2}{*}{$\Big\{E\Big\}$}&2\,$\omega$&$\chi_{01}$&1.9566\,$\omega$&$\chi_{01}$&$1.35\times 10^{-4}$\\
2&&2\,$\omega$&$\chi_{10}$&1.9566\,$\omega$&$\chi_{10}$&$1.35\times 10^{-4}$\\
3&$B_2$&3\,$\omega$&$\chi_{11}$&2.8665\,$\omega$&$\chi_{11}$&$3.38\times 10^{-3}$\\
4&$A_1$&3\,$\omega$&$(\chi_{02}+\chi_{20})/\sqrt{2}$&2.8942\,$\omega$&$(\chi_{02}+\chi_{20})/\sqrt{2}$&$3.71\times 10^{-3}$\\
5&$B_1$&3\,$\omega$&$(\chi_{02}-\chi_{20})/\sqrt{2}$&2.9572\,$\omega$&$(\chi_{02}-\chi_{20})/\sqrt{2}$&$4.60\times 10^{-3}$\\
6&\multirow{2}{*}{$\Big\{E\Big\}$}&4\,$\omega$&$\chi_{12}$&3.7490\,$\omega$&$0.937\chi_{12}+0.349\chi_{30}$&$6.44\times 10^{-2}$\\
7&&4\,$\omega$&$\chi_{21}$&3.7490\,$\omega$&$0.937\chi_{21}+0.349\chi_{03}$&$6.44\times 10^{-2}$\\
8&\multirow{2}{*}{$\Big\{E\Big\}$}&4\,$\omega$&$\chi_{03}$&3.9156\,$\omega$&$0.937\chi_{30}-0.349\chi_{12}$&$0.11$\\
9&&4\,$\omega$&$\chi_{30}$&3.9156\,$\omega$&$0.937\chi_{03}-0.349\chi_{21}$&$0.11$\\
10&$A_1$&5\,$\omega$&$\chi_{22}$&4.5616\,$\omega$&$0.937\chi_{22}+0.257(\chi_{40}+\chi_{04})$&$0.87$\\
11&$B_2$&5\,$\omega$&$(\chi_{13}+\chi_{31})/\sqrt{2}$&4.5684\,$\omega$&$(\chi_{13}+\chi_{31})/\sqrt{2}$&$0.89$\\
12&$A_2$&5\,$\omega$&$(\chi_{13}-\chi_{31})/\sqrt{2}$&$>\Delta E_\kappa$
&$(\chi_{13}-\chi_{31})/\sqrt{2}$&--\\
12&$B_1$&5\,$\omega$&$(\chi_{04}-\chi_{40})/\sqrt{2}$&$>\Delta E_\kappa$
&$(\chi_{04}-\chi_{40})/\sqrt{2}$&--\\
12&$A_1$&5\,$\omega$&$(\chi_{04}+\chi_{40})/\sqrt{2}$&$>\Delta E_\kappa$
&$0.663(\chi_{04}+\chi_{40})-0.349\chi_{22}$&--
\end{tabular}
\end{ruledtabular}
\end{table*}

The energies $E_{\kappa,i}^{cm}$ of the first 12 eigenstates are tabulated
in Tab.~\ref{tab:cmstates} for both potentials
$E_\kappa^{(0)}(\mathbf R)$ (``h.o.'', third column) and
$E_\kappa^{(0)}(\mathbf R)+E_\kappa^{(2)}(\mathbf R)$
(``perturbed'', fifth column).
While $E_\kappa^{(0)}(\mathbf R)$ yields energies
$E_\nu=(\nu+1)\omega$ with $\nu=0,1,2,\dots$
(we assumed a perfectly harmonic potential), the eigenenergies
belonging to $E_\kappa^{(0)}(\mathbf R)+E_\kappa^{(2)}(\mathbf R)$
deviate from this rule:
as the harmonic potential is flattened by the contribution
$E_\kappa^{(2)}(\mathbf R)$, the energies are below the harmonic ones.
The remaining degeneracies which appear for odd $\nu$ (e.g., states 6--9)
can be explained by the symmetry properties of the involved states:
in this case, only $E$ symmetry is encountered.
The latter has a two-dimensional irreducible representation
hence the appearance of $\frac{\nu+1}{2}$ degenerate pairs.

Finally, let us briefly comment on the issue of tunneling.
States that are confined within the potentials shown in
Sec.~\ref{sec:surfaces} may escape the trap by tunneling
through the potential barrier along the diagonals.
While this process most certainly plays no role
for configurations where the time scale of the trap frequency
is large compared to the radiative lifetime,
\emph{a priori} it is not clear if tunneling becomes crucial for
tighter traps.
For this reason, we estimated the lifetime associated
with the tunneling process by investigating the
transmission probability for the $i$th excited c.m.\ state
in one dimension,
$P_t=\exp(-2\int_a^b\sqrt{2M[E_\kappa(X=Y)-E_{\kappa,i}^{cm}]}\,\mathrm dX\,)$,
where the integration limits $a$ and $b$ are determined by
the condition $E_\kappa(X=Y)=E_{\kappa,i}^{cm}$.
Since the Rydberg atom `hits' the potential barriers twice per trapping
period, the loss rate can be roughly estimated by $2\omega P_t$.
Actual values of $P_t$ for the c.m.\ states discussed in this
section are given in the last column of Tab.~\ref{tab:cmstates}.
For tighter magnetic traps, where more c.m.\ states can
be confined, $P_t$ substantially decreases further.
Hence, tunneling only has to be considered for highly
excited c.m.\ excitations close to the top of the barrier.

\section{Parametric Heating}
\label{sec:heating}
Utilizing state-dependent (Rydberg-Rydberg) interactions
for quantum information protocols
necessitates the excitation of trapped ground state atoms
to a Rydberg state by a $\pi$-pulse 
\cite{PhysRevLett.85.2208,PhysRevLett.87.037901}.
When the excitation process is much shorter than the timescale of the 
external motion, such an excitation effectively causes a sudden change
of the trapping potential.
This couples and thus redistributes the initial c.m.\ quantum state
to neighboring levels which, in general, increases 
the c.m.\ energy (hence we will denote
this process as ``parametric heating'' in the following).
In this section, we investigate this effect and calculate the
corresponding heating rates.

Suppose we have a $^{87}$Rb atom in its $5S_{1/2}, F=m_F=2$
electronic ground state which is at $t=0$ instantaneously
excited to the Rydberg state $40S_{1/2}, m_j=1/2$ and after a
short period of time $t'$ again de-excited to its electronic
ground state.
Furthermore, we assume the atom to reside in a well defined c.m.\
state at $t=0$, i.e., $\chi(\mathbf R,t=0)=\chi_{\nu_x\nu_y}(\mathbf R)$;
note that $\chi_{\nu_x\nu_y}(\mathbf R)$ denote the c.m.\
eigenfunctions of the ground state atom rather than
the Rydberg atom.
Except for the contribution $E_\kappa^{(2)}(\mathbf R)$,
both electronic states give rise to the same
trapping potential $E_\kappa^{(0)}(\mathbf R)$
\footnote{We remark that for the Rydberg state
the hyperfine interaction can be treated perturbatively and does not
alter the trapping potentials for the regime of field strengths
we are considering.},
i.e., in the simplest approximation [which is neglecting
$E_\kappa^{(2)}(\mathbf R)$]
the c.m.\ state is not affected by the excitation to the
Rydberg level.
If we account for the extra term $E_\kappa^{(2)}(\mathbf R)$,
on the other hand, the situation changes substantially.
We consider the sequence \textit{ground state $\rightarrow$ Rydberg
state $\rightarrow$ ground state}, where all transitions are carried 
out by fast $\pi$-pulses. $E_\kappa^{(2)}(\mathbf R)$ can then be 
considered as a perturbation of the ground state trapping potential 
which acts for the time interval during which the atom resides in the 
Rydberg level, i.e., $0<t<t'$.
As shown in Sec.~\ref{sec:cmwave}, $E_\kappa^{(2)}(\mathbf R)$
mixes c.m.\ states according to the selection rules $\Delta \nu_{x/y}=0$ 
and $\Delta \nu_{x/y}=\pm2$;
hence the Rydberg excitation leads for the ground state atom
to the admixture of  lower- and higher-lying c.m.\
levels with $\nu'=\nu$, $\nu'=\nu\pm2$, and $\nu'=\nu\pm4$,
where $\nu=\nu_x+\nu_y$. Note that we adopt here again the approximation
of a purely harmonic potential $E_\kappa^{(0)}(\mathbf R)\propto\frac{1}{2}M\omega^2(X^2+Y^2)$.

Within time-dependent perturbation theory, the probability
of a transition $|\nu_x\nu_y\rangle\rightarrow|\nu_x'\nu_y'\rangle$
of the c.m.\ state of a ground state atom
due to its short-time Rydberg excitation is consequently given by
\begin{equation}
 W_{\nu_x\nu_y\rightarrow\nu_x'\nu_y'}=
\left|\langle\nu_x'\nu_y'|E_\kappa^{(2)}(\mathbf R)|\nu_x\nu_y\rangle\right|^2
 f(t',\tilde\omega)
\end{equation}
with
$f(t',\tilde\omega)=\frac{\sin^2(\tilde\omega t'/2)}{(\tilde\omega/2)^2}$
and $\tilde\omega=(\nu'-\nu)\, \omega=\Delta\nu\cdot \omega$
\cite{friedrich98}.
The average rate to make a transition to state $|\nu_x'\nu_y'\rangle$
within the time interval $t'$ consequently reads
\begin{equation}
 R_{\nu_x\nu_y\rightarrow\nu_x'\nu_y'}=\frac{1}{t'}
W_{\nu_x\nu_y\rightarrow\nu_x'\nu_y'}\,.
\end{equation}
This allows us to define a heating rate as
\begin{align}
 \dot E_{\nu_x\nu_y}={}&\sum_{\nu_x'\nu_y'}\tilde
\omega R_{\nu_x\nu_y\rightarrow\nu_x'\nu_y'}\\
={}&\Big\{
2\frac{\sin^2(\omega t')}{\omega^2}(\nu_x+\nu_y+1)
(\nu_x+\frac{1}{2})(\nu_y+\frac{1}{2})+\nonumber\\
&\frac{\sin^2(2\omega t')}{4\omega^2}(\nu_x+\nu_y+1)\Big[
(\nu_x+\frac{1}{2})(\nu_y+\frac{1}{2})+\frac{3}{4}\Big]\Big\}
\nonumber\\
&\times\frac{C_z^2G^4}{M^4\omega^4}\frac{\omega}{t'}
\label{eq:heating}
\end{align}
where we used the recurrence relation
$\langle \nu_x'|X^2|\nu_x\rangle=\frac{1}{2M\omega}
\big[\sqrt{\nu_x(\nu_x-1)}\delta_{\nu_x'\nu_x-2}+
(2\nu_x+1)\delta_{\nu_x'\nu_x}+
\sqrt{(\nu_x+1)(\nu_x+2)}\delta_{\nu_x'\nu_x+2}\big]$ of
the harmonic oscillator eigenfunctions
and assumed $E_\kappa^{(2)}(\mathbf R)=C_zG^2X^2Y^2$.
Note that $\dot E_{\nu_x\nu_y}>0$ independent of the initial state,
i.e., cooling is not possible.
For short times $t'\ll 1/\omega$, one can approximate
$\frac{\sin^2(\omega t')}{\omega^2}\approx t'^2$
which gives an overall linear increase of the heating rate
in time.

\begin{figure}
\includegraphics[width=8.0cm]{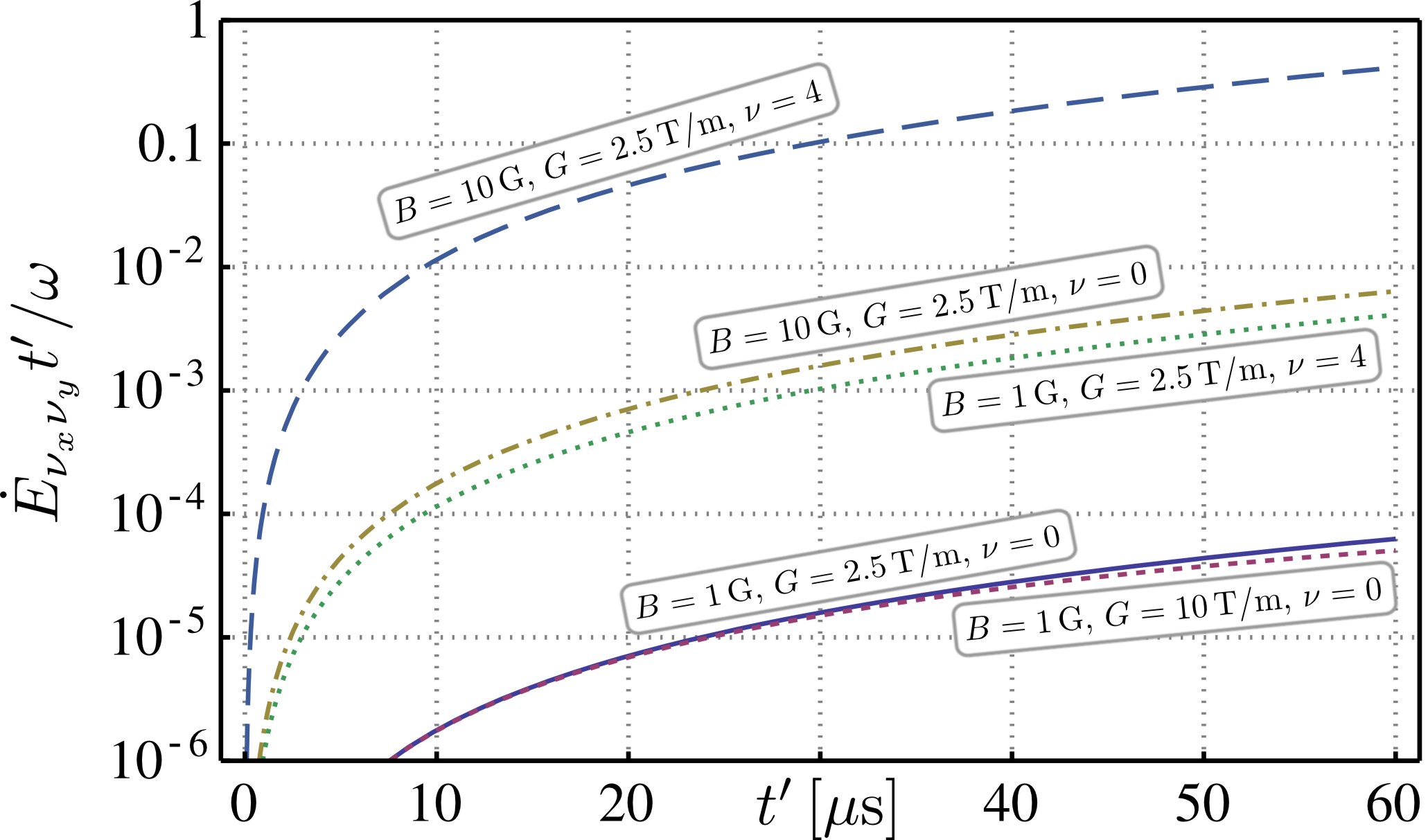}
\caption{(Color online) Parametric heating $\dot E_{\nu_x\nu_y} t'/\omega$
of a trapped ground state atom
as a function of the time being excited to the Rydberg level $40S_{1/2}$.
Several initial c.m.\ states and magnetic field configurations are
considered:
$B=1\,$G, $G=2.5\,$T/m (solid),
$B=1\,$G, $G=10\,$T/m (short-dashed), and
$B=10\,$G, $G=2.5\,$T/m (dashed-dotted)
for $\nu_x=\nu_y=0$ as well as
$B=1\,$G, $G=2.5\,$T/m (dotted) and
$B=10\,$G, $G=2.5\,$T/m (long-dashed)
for $\nu_x=\nu_y=2$.}
\label{fig:heating}
\end{figure}

In Figure \ref{fig:heating}, the parametric heating
$\dot E_{\nu_x\nu_y} t'$
in terms of the trap frequency $\omega$ and
as a function of the Rydberg excitation period $t'$ is
illustrated for several c.m.\ initial states and magnetic field
configurations for the Rydberg state $40S_{1/2}$.
As one can observe, the heating mainly depends on the
Ioffe field strength $B$ rather than on the magnetic field
gradient $G$. An increase of the latter barely changes
$\dot E_{\nu_x\nu_y} t'/\omega$ while a stronger Ioffe field
results in a substantial increase.
As expected from Eq.~(\ref{eq:heating}), $\dot E_{\nu_x\nu_y} t'/\omega$
also significantly increases if the ground state atom is
initially in an excited c.m.\ state.
However, for the given examples the overall
heating within
the radiative lifetime of the Rydberg atom turns out to be
very moderate with $\dot E_{\nu_x\nu_y} t'<1\,\omega$
(for $B=1$ G and $G=2.5$ T/m, $\dot E_{\nu_x\nu_y} t'=1\,\omega$
corresponds to 15 nK).
Hence, only for high c.m.\ levels $\nu=\nu_x+\nu_y$
and long times $t'$ the above described excitation of the c.m.\ motion
of an ultracold sample of Rb atoms due to the Rydberg excitation
is expected to become an issue.

Finally, let us briefly comment on what is expected for a thermal atom
where the c.m.\ state is not a pure state but rather a mixture according
to the Boltzmann distribution
$f_\nu(T)=g_\nu e^{-(\nu+1)\omega/k_bT}/Z(T)$,
$Z(T)=\sum_{\nu=0}^\infty g_\nu e^{-(\nu+1)\omega/k_bT}$ being the partition
function and $g_\nu=\nu+1$ the degeneracy of the $\nu$th excited
c.m.\ state.
In this case, the heating rate reads
\begin{align}
 \dot E(T)&=\sum_{\nu=0}^\infty f_\nu \dot E_{\nu_x\nu_y}\\
&\approx
\frac{3}{2}\frac{C_z^2G^4}{M^4\omega^3}t'\coth^3\!
\left(\frac{1}{2}\frac{\omega}{k_bT}\right)
\label{eq:thermale1}\\
&\approx\frac{12C_z^2G^4}{M^4\omega^3}t'
\Big(\frac{k_bT}{\omega}\Big)^3\,.
\label{eq:thermale2}
\end{align}
Equation (\ref{eq:thermale1}) is obtained by approximating
$\frac{\sin^2(\omega t')}{\omega^2}\approx t'^2$
for short times $t'\ll 1/\omega$ and further
simplified to Eq.~(\ref{eq:thermale2}) by assuming $k_bT\gg \omega$.
As expected from Eq.~(\ref{eq:heating}), $\dot E(T)$
rapidly increases with the temperature $T$ since higher c.m.\
excitations are populated.

\section{Dephasing}
\label{sec:dephasing}
Besides the parametric heating due to the short-time Rydberg excitation
of a ground state atom -- as discussed in the previous section --
the dephasing of the c.m.\ motion of the Rydberg and the ground state
might become an issue for experimental schemes realizing quantum
information protocols.
Let us consider the situation as described in Ref.\
\cite{PhysRevLett.85.2208},
i.e., we have two ground states denoted by $|0\rangle$ and $|1\rangle$
where only the latter is coupled to a Rydberg state $|r\rangle$
by a laser transition.
The density operator of the internal degree of freedom, i.e., only
considering the electronic state, of such a two state system can
generally be written as
$\rho_\mathrm{int}=
a|0\rangle\langle0|+(1-a)|1\rangle\langle1|
+b|1\rangle\langle0|+b^*|0\rangle\langle1|$
giving rise to the density matrix
\begin{equation}
 \rho_\mathrm{int}=\begin{pmatrix}
       a&b^*\\
       b&1-a
      \end{pmatrix}\,.
\label{eq:intmatrix}
\end{equation}
If we assume furthermore that both ground states are
identically prepared with respect to their external, i.e., c.m.\
motion, the total density matrix factorizes into an internal
and external contribution,
$\rho_\mathrm{tot}=\rho_\mathrm{int}\otimes\rho_\mathrm{ext}$,
where
$\rho_\mathrm{ext}=\sum_{\nu_y\nu_y}p_{\nu_x\nu_y}
|\nu_x\nu_y\rangle\langle\nu_x\nu_y|$.

For various implementations of quantum information protocols now
the Rydberg state $|r\rangle$ comes into play.
Suppose that state $|1\rangle$ is excited to $|r\rangle$ for a
given time $t'$.
As pointed out in Sec.~\ref{sec:heating}, this will influence its
c.m.\ motion by causing transitions
$|\nu_x\nu_y\rangle\rightarrow
|\tilde\chi\rangle$, where
$|\tilde\chi\rangle\equiv\sum_{\nu_x'\nu_y'}
C_{\nu_x\nu_y}^{\nu_x'\nu_y'}(t')|\nu_x'\nu_y'\rangle$;
$C_{\nu_x\nu_y}^{\nu_x'\nu_y'}(t')$ denotes the amplitude for
being at time $t'$ in state $|\nu_x'\nu_y'\rangle$ if initially
residing in state $|\nu_x\nu_y\rangle$.
Hence, after the short-time Rydberg excitation of solely state
$|1\rangle$, the density matrix does not decouple anymore and
consequently reads
\begin{align}
 \rho_\mathrm{tot}={}&\sum_{\nu_x\nu_y}p_{\nu_x\nu_y}
\big[a|0\rangle\langle0|\otimes|\nu_x\nu_y\rangle\langle\nu_x\nu_y|
\nonumber\\
&\qquad+(1-a)|1\rangle\langle1|\otimes|\tilde\chi\rangle\langle\tilde\chi|
\nonumber\\
&\qquad+b|1\rangle\langle0|\otimes|\tilde\chi\rangle\langle\nu_x\nu_y|
\nonumber\\
&\qquad+b^*|0\rangle\langle1|\otimes|\nu_x\nu_y\rangle\langle\tilde\chi|\big].
\end{align}
Any qubit-related measurement, however, only acts on the internal
degrees of freedom, i.e., the electronic states.
As a consequence, the relevant object in this case is the
reduced density matrix where the c.m.\ degree of freedom
is traced out.
Defining $\beta=\sum_{\nu_x\nu_y}p_{\nu_x\nu_y}
\langle\nu_x\nu_y|\tilde\chi\rangle$, one eventually yields
\begin{equation}
 \mathrm{Tr}_\mathrm{ext}\rho_\mathrm{tot}=\begin{pmatrix}
       a&b^*\beta^*\\
       b\beta&1-a
      \end{pmatrix}\,.
\end{equation}
Comparing this result to the case where the internal and external
degree of freedom factorize,
$\mathrm{Tr}_\mathrm{ext}(\rho_\mathrm{int}\otimes\rho_\mathrm{ext})
=\rho_\mathrm{int}$, it is clear that the short-time Rydberg excitation
will inevitably influence the properties of our system.
In order to quantify this effect, we consider the purity
$\mathcal P(\rho)=\mathrm{Tr}\rho^2$
of the reduced density matrix.
In particular, if the system is initially prepared in a pure internal
state (as for example $(|0\rangle\pm|1\rangle)/\sqrt{2}$,
which is envisaged for the realization of a two
qubit \textsc{cnot} gate \cite{Urban2009})
the above described process is expected to decrease its purity.
Indeed, one yields
\begin{align}
 \mathcal P(\mathrm{Tr}_\mathrm{ext}\rho_\mathrm{tot})&=
a^2+(1-a)^2+2|b|^2|\beta|^2\\
&=\mathcal P_\mathrm{int}-2|b|^2(1-|\beta|^2)
\end{align}
where $\mathcal P_\mathrm{int}=a^2+(1-a)^2+2|b|^2$ denotes the purity
of the system only considering the internal degree of freedom.
Note that for any pure state
$\mathcal P_\mathrm{int}=1$ is found.
Hence, the reduction of the purity is determined by the magnitude of
$|\beta|^2$ and therefore by the overlap integrals
$\langle\nu_x\nu_y|\tilde \chi\rangle$ of the c.m.\ wavefunction
after the Rydberg excitation.

A particularly illustrative situation arises, if the atom is initially
prepared in its c.m.\ ground state, i.e.,
$p_{\nu_x\nu_y}=\delta_{0\nu_x}\delta_{0\nu_y}$ and
$|\beta|^2=|\langle00|\tilde\chi\rangle|^2$ correspondingly.
In this case, $|\beta|^2$ is given by the probability of finding
the atom still in the c.m.\ ground state after being excited to the
Rydberg state for the time $t'$.
According to Sec.~\ref{sec:heating}, we find
\begin{align}
 |\beta|^2&=1-\sum_{\nu_x\nu_y\neq00}
W_{00\rightarrow\nu_x\nu_y}\\
&=1-\frac{C_z^2G^4}{4M^4\omega^4}
\left(\frac{\sin^2(\omega t')}{\omega^2}+
\frac{\sin^2(2\omega t')}{4\omega^2}\right)\\
&\approx1-\frac{C_z^2G^4}{2M^4\omega^4}{t'}^2
\end{align}
and therefore
\begin{equation}
 \mathcal P(\mathrm{Tr}_\mathrm{ext}\rho_\mathrm{tot})\approx
 \mathcal P_\mathrm{int}-2|b|^2\frac{C_z^2G^4}{2M^4\omega^4}{t'}^2\,.
\end{equation}
As expected, the decrease of the purity depends explicitly
(for short times $t'$ even quadratically) on the
time $t'$ of being excited to the Rydberg level.

\section{Conclusion}
We theoretically investigated the quantum properties of
Rydberg atoms in a magnetic Ioffe-Pritchard trap. 
In particular, the electronic properties and the center of mass dynamics of
the low angular momentum $nS$, $nP$, and $nD$ states of $^{87}$Rb have been studied.
It turns out that the composite nature of Rydberg atoms,
i.e., the fact that it consists of an outer electron far away
from a compact ionic core, significantly
alters the coupling of the electronic motion to the inhomogeneous
magnetic field of the Ioffe-Pritchard trap.
We demonstrated that this leads to qualitative changes in the
trapping potentials, namely, the appearance of an de-confining
contribution which reduces the azimuthal symmetry to ${\bm C}_{4v}$.
As a consequence, the resulting energy surfaces -- which characterize
the trapping potentials -- possess a finite depth.
Analytical expressions describing the surfaces were derived
and the applicability of the applied perturbative treatment has been verified
for experimentally relevant field strengths
by comparison with numerical solutions of the underlying Schr\"odinger equation.
Exemplary energy surfaces of the fully polarized $n=40$, $l=0,1,2$, $m_j=j$ states
for the magnetic field configuration $B=1$ G, $G=2.5$ T/m were provided.
A clear deviation from the harmonic confinement of a point-like particle with
a trap depth of only a few vibrational quanta could be observed.
Choosing different magnetic field parameters, on the other hand,
trapping can be achieved with trap depths in the micro-Kelvin regime.
The non-harmonicity of the Rydberg trapping potential
becomes also apparent in the resulting center of mass dynamics:
The additional contribution due to the two-body character of the Rydberg atom
mixes the ``unperturbed'' harmonic eigenstates and thereby partially
lifts their degeneracy.
For an atom in its electronic ground state that is
excited to a Rydberg state only for a short
period of time, this provides a mechanism 
for parametric heating by populating excited center of mass states.
The corresponding heating rate as a function of the initial
center of mass state of the ground state atom has been derived.
In the framework of quantum information protocols involving
the short-time population of Rydberg atoms,
it has been demonstrated that the same mechanism
can lead to a decrease of the purity of the involved
qubit states.

A rather natural extension of the present work would be
the investigation of magnetic field configurations other
than the Ioffe-Pritchard trap.
For example, it is expected also in the case of a three-dimensional
quadrupole field that similar terms associated with the composite nature
of the Rydberg atom arise, significantly altering the trapping
potential compared to the point-like particle description.

\begin{acknowledgments}
This work was supported by the German Research Foundation (DFG)
within the framework of the
Excellence Initiative through the Heidelberg Graduate School of
Fundamental Physics (Grant No.\ GSC~129/1).
M.M.\ acknowledges financial support from the Landesgraduiertenf\"orderung
Baden-W\"urttemberg.
Financial support by the DFG through Grant No.\ Schm 885/10-3
is gratefully acknowledged.
\end{acknowledgments}

\bibliography{rydsurf}

\begin{thebibliography}{37}
\expandafter\ifx\csname natexlab\endcsname\relax\def\natexlab#1{#1}\fi
\expandafter\ifx\csname bibnamefont\endcsname\relax
  \def\bibnamefont#1{#1}\fi
\expandafter\ifx\csname bibfnamefont\endcsname\relax
  \def\bibfnamefont#1{#1}\fi
\expandafter\ifx\csname citenamefont\endcsname\relax
  \def\citenamefont#1{#1}\fi
\expandafter\ifx\csname url\endcsname\relax
  \def\url#1{\texttt{#1}}\fi
\expandafter\ifx\csname urlprefix\endcsname\relax\def\urlprefix{URL }\fi
\providecommand{\bibinfo}[2]{#2}
\providecommand{\eprint}[2][]{\url{#2}}

\bibitem[{\citenamefont{Gallagher}(1994)}]{gallagher94}
\bibinfo{author}{\bibfnamefont{T.~F.} \bibnamefont{Gallagher}},
  \emph{\bibinfo{title}{Rydberg Atoms}} (\bibinfo{publisher}{Cambridge
  University Press}, \bibinfo{address}{Cambridge, U.K.}, \bibinfo{year}{1994}).

\bibitem[{\citenamefont{Jaksch et~al.}(2000)\citenamefont{Jaksch, Cirac,
  Zoller, Rolston, C\^ot\'e, and Lukin}}]{PhysRevLett.85.2208}
\bibinfo{author}{\bibfnamefont{D.}~\bibnamefont{Jaksch}},
  \bibinfo{author}{\bibfnamefont{J.~I.} \bibnamefont{Cirac}},
  \bibinfo{author}{\bibfnamefont{P.}~\bibnamefont{Zoller}},
  \bibinfo{author}{\bibfnamefont{S.~L.} \bibnamefont{Rolston}},
  \bibinfo{author}{\bibfnamefont{R.}~\bibnamefont{C\^ot\'e}}, \bibnamefont{and}
  \bibinfo{author}{\bibfnamefont{M.~D.} \bibnamefont{Lukin}},
  \bibinfo{journal}{Phys. Rev. Lett.} \textbf{\bibinfo{volume}{85}},
  \bibinfo{pages}{2208} (\bibinfo{year}{2000}).

\bibitem[{\citenamefont{Lukin et~al.}(2001)\citenamefont{Lukin, Fleischhauer,
  C\^ot\'e, Duan, Jaksch, Cirac, and Zoller}}]{PhysRevLett.87.037901}
\bibinfo{author}{\bibfnamefont{M.~D.} \bibnamefont{Lukin}},
  \bibinfo{author}{\bibfnamefont{M.}~\bibnamefont{Fleischhauer}},
  \bibinfo{author}{\bibfnamefont{R.}~\bibnamefont{C\^ot\'e}},
  \bibinfo{author}{\bibfnamefont{L.~M.} \bibnamefont{Duan}},
  \bibinfo{author}{\bibfnamefont{D.}~\bibnamefont{Jaksch}},
  \bibinfo{author}{\bibfnamefont{J.~I.} \bibnamefont{Cirac}}, \bibnamefont{and}
  \bibinfo{author}{\bibfnamefont{P.}~\bibnamefont{Zoller}},
  \bibinfo{journal}{Phys. Rev. Lett.} \textbf{\bibinfo{volume}{87}},
  \bibinfo{pages}{037901} (\bibinfo{year}{2001}).

\bibitem[{\citenamefont{Tong et~al.}(2004)\citenamefont{Tong, Farooqi,
  Stanojevic, Krishnan, Zhang, C\^ot\'e, Eyler, and Gould}}]{tong:063001}
\bibinfo{author}{\bibfnamefont{D.}~\bibnamefont{Tong}},
  \bibinfo{author}{\bibfnamefont{S.~M.} \bibnamefont{Farooqi}},
  \bibinfo{author}{\bibfnamefont{J.}~\bibnamefont{Stanojevic}},
  \bibinfo{author}{\bibfnamefont{S.}~\bibnamefont{Krishnan}},
  \bibinfo{author}{\bibfnamefont{Y.~P.} \bibnamefont{Zhang}},
  \bibinfo{author}{\bibfnamefont{R.}~\bibnamefont{C\^ot\'e}},
  \bibinfo{author}{\bibfnamefont{E.~E.} \bibnamefont{Eyler}}, \bibnamefont{and}
  \bibinfo{author}{\bibfnamefont{P.~L.} \bibnamefont{Gould}},
  \bibinfo{journal}{Phys. Rev. Lett.} \textbf{\bibinfo{volume}{93}},
  \bibinfo{pages}{063001} (\bibinfo{year}{2004}).

\bibitem[{\citenamefont{Singer et~al.}(2004)\citenamefont{Singer, Reetz-Lamour,
  Amthor, Marcassa, and Weidem\"uller}}]{singer:163001}
\bibinfo{author}{\bibfnamefont{K.}~\bibnamefont{Singer}},
  \bibinfo{author}{\bibfnamefont{M.}~\bibnamefont{Reetz-Lamour}},
  \bibinfo{author}{\bibfnamefont{T.}~\bibnamefont{Amthor}},
  \bibinfo{author}{\bibfnamefont{L.~G.} \bibnamefont{Marcassa}},
  \bibnamefont{and}
  \bibinfo{author}{\bibfnamefont{M.}~\bibnamefont{Weidem\"uller}},
  \bibinfo{journal}{Phys. Rev. Lett.} \textbf{\bibinfo{volume}{93}},
  \bibinfo{pages}{163001} (\bibinfo{year}{2004}).

\bibitem[{\citenamefont{{Cubel Liebisch} et~al.}(2005)\citenamefont{{Cubel
  Liebisch}, Reinhard, Berman, and Raithel}}]{liebisch:253002}
\bibinfo{author}{\bibfnamefont{T.}~\bibnamefont{{Cubel Liebisch}}},
  \bibinfo{author}{\bibfnamefont{A.}~\bibnamefont{Reinhard}},
  \bibinfo{author}{\bibfnamefont{P.~R.} \bibnamefont{Berman}},
  \bibnamefont{and} \bibinfo{author}{\bibfnamefont{G.}~\bibnamefont{Raithel}},
  \bibinfo{journal}{Phys. Rev. Lett.} \textbf{\bibinfo{volume}{95}},
  \bibinfo{pages}{253002} (\bibinfo{year}{2005}).

\bibitem[{\citenamefont{Vogt et~al.}(2007)\citenamefont{Vogt, Viteau, Chotia,
  Zhao, Comparat, and Pillet}}]{vogt:073002}
\bibinfo{author}{\bibfnamefont{T.}~\bibnamefont{Vogt}},
  \bibinfo{author}{\bibfnamefont{M.}~\bibnamefont{Viteau}},
  \bibinfo{author}{\bibfnamefont{A.}~\bibnamefont{Chotia}},
  \bibinfo{author}{\bibfnamefont{J.}~\bibnamefont{Zhao}},
  \bibinfo{author}{\bibfnamefont{D.}~\bibnamefont{Comparat}}, \bibnamefont{and}
  \bibinfo{author}{\bibfnamefont{P.}~\bibnamefont{Pillet}},
  \bibinfo{journal}{Phys. Rev. Lett.} \textbf{\bibinfo{volume}{99}},
  \bibinfo{pages}{073002} (\bibinfo{year}{2007}).

\bibitem[{\citenamefont{van Ditzhuijzen et~al.}(2008)\citenamefont{van
  Ditzhuijzen, Koenderink, Hern\'{a}ndez, Robicheaux, Noordam, and van Linden
  van~den Heuvell}}]{ditzhuijzen:243201}
\bibinfo{author}{\bibfnamefont{C.~S.~E.} \bibnamefont{van Ditzhuijzen}},
  \bibinfo{author}{\bibfnamefont{A.~F.} \bibnamefont{Koenderink}},
  \bibinfo{author}{\bibfnamefont{J.~V.} \bibnamefont{Hern\'{a}ndez}},
  \bibinfo{author}{\bibfnamefont{F.}~\bibnamefont{Robicheaux}},
  \bibinfo{author}{\bibfnamefont{L.~D.} \bibnamefont{Noordam}},
  \bibnamefont{and} \bibinfo{author}{\bibfnamefont{H.~B.} \bibnamefont{van
  Linden van~den Heuvell}}, \bibinfo{journal}{Phys. Rev. Lett.}
  \textbf{\bibinfo{volume}{100}}, \bibinfo{pages}{243201}
  (\bibinfo{year}{2008}).

\bibitem[{\citenamefont{Heidemann et~al.}(2007)\citenamefont{Heidemann,
  Raitzsch, Bendkowsky, Butscher, L\"{o}w, Santos, and
  Pfau}}]{heidemann:163601}
\bibinfo{author}{\bibfnamefont{R.}~\bibnamefont{Heidemann}},
  \bibinfo{author}{\bibfnamefont{U.}~\bibnamefont{Raitzsch}},
  \bibinfo{author}{\bibfnamefont{V.}~\bibnamefont{Bendkowsky}},
  \bibinfo{author}{\bibfnamefont{B.}~\bibnamefont{Butscher}},
  \bibinfo{author}{\bibfnamefont{R.}~\bibnamefont{L\"{o}w}},
  \bibinfo{author}{\bibfnamefont{L.}~\bibnamefont{Santos}}, \bibnamefont{and}
  \bibinfo{author}{\bibfnamefont{T.}~\bibnamefont{Pfau}},
  \bibinfo{journal}{Phys. Rev. Lett.} \textbf{\bibinfo{volume}{99}},
  \bibinfo{pages}{163601} (\bibinfo{year}{2007}).

\bibitem[{\citenamefont{Reetz-Lamour et~al.}(2008)\citenamefont{Reetz-Lamour,
  Amthor, Deiglmayr, and Weidem\"{u}ller}}]{reetz-lamour:253001}
\bibinfo{author}{\bibfnamefont{M.}~\bibnamefont{Reetz-Lamour}},
  \bibinfo{author}{\bibfnamefont{T.}~\bibnamefont{Amthor}},
  \bibinfo{author}{\bibfnamefont{J.}~\bibnamefont{Deiglmayr}},
  \bibnamefont{and}
  \bibinfo{author}{\bibfnamefont{M.}~\bibnamefont{Weidem\"{u}ller}},
  \bibinfo{journal}{Phys. Rev. Lett.} \textbf{\bibinfo{volume}{100}},
  \bibinfo{pages}{253001} (\bibinfo{year}{2008}).

\bibitem[{\citenamefont{Johnson et~al.}(2008)\citenamefont{Johnson, Urban,
  Henage, Isenhower, Yavuz, Walker, and Saffman}}]{johnson:113003}
\bibinfo{author}{\bibfnamefont{T.~A.} \bibnamefont{Johnson}},
  \bibinfo{author}{\bibfnamefont{E.}~\bibnamefont{Urban}},
  \bibinfo{author}{\bibfnamefont{T.}~\bibnamefont{Henage}},
  \bibinfo{author}{\bibfnamefont{L.}~\bibnamefont{Isenhower}},
  \bibinfo{author}{\bibfnamefont{D.~D.} \bibnamefont{Yavuz}},
  \bibinfo{author}{\bibfnamefont{T.~G.} \bibnamefont{Walker}},
  \bibnamefont{and} \bibinfo{author}{\bibfnamefont{M.}~\bibnamefont{Saffman}},
  \bibinfo{journal}{Phys. Rev. Lett.} \textbf{\bibinfo{volume}{100}},
  \bibinfo{pages}{113003} (\bibinfo{year}{2008}).

\bibitem[{\citenamefont{Urban et~al.}(2009)\citenamefont{Urban, Johnson,
  Henage, Isenhower, Yavuz, Walker, and Saffman}}]{Urban2009}
\bibinfo{author}{\bibfnamefont{E.}~\bibnamefont{Urban}},
  \bibinfo{author}{\bibfnamefont{T.~A.} \bibnamefont{Johnson}},
  \bibinfo{author}{\bibfnamefont{T.}~\bibnamefont{Henage}},
  \bibinfo{author}{\bibfnamefont{L.}~\bibnamefont{Isenhower}},
  \bibinfo{author}{\bibfnamefont{D.~D.} \bibnamefont{Yavuz}},
  \bibinfo{author}{\bibfnamefont{T.~G.} \bibnamefont{Walker}},
  \bibnamefont{and} \bibinfo{author}{\bibfnamefont{M.}~\bibnamefont{Saffman}},
  \bibinfo{journal}{Nat. Phys.} \textbf{\bibinfo{volume}{5}},
  \bibinfo{pages}{110} (\bibinfo{year}{2009}).

\bibitem[{\citenamefont{Ga\"etan et~al.}(2009)\citenamefont{Ga\"etan,
  Miroshnychenko, Wilk, Chotia, Viteau, Comparat, Pillet, Browaeys, and
  Grangier}}]{Gaetan2009}
\bibinfo{author}{\bibfnamefont{A.}~\bibnamefont{Ga\"etan}},
  \bibinfo{author}{\bibfnamefont{Y.}~\bibnamefont{Miroshnychenko}},
  \bibinfo{author}{\bibfnamefont{T.}~\bibnamefont{Wilk}},
  \bibinfo{author}{\bibfnamefont{A.}~\bibnamefont{Chotia}},
  \bibinfo{author}{\bibfnamefont{M.}~\bibnamefont{Viteau}},
  \bibinfo{author}{\bibfnamefont{D.}~\bibnamefont{Comparat}},
  \bibinfo{author}{\bibfnamefont{P.}~\bibnamefont{Pillet}},
  \bibinfo{author}{\bibfnamefont{A.}~\bibnamefont{Browaeys}}, \bibnamefont{and}
  \bibinfo{author}{\bibfnamefont{P.}~\bibnamefont{Grangier}},
  \bibinfo{journal}{Nat. Phys.} \textbf{\bibinfo{volume}{5}},
  \bibinfo{pages}{115} (\bibinfo{year}{2009}).

\bibitem[{\citenamefont{Hyafil et~al.}(2004)\citenamefont{Hyafil, Mozley,
  Perrin, Tailleur, Nogues, Brune, Raimond, and Haroche}}]{hyafil:103001}
\bibinfo{author}{\bibfnamefont{P.}~\bibnamefont{Hyafil}},
  \bibinfo{author}{\bibfnamefont{J.}~\bibnamefont{Mozley}},
  \bibinfo{author}{\bibfnamefont{A.}~\bibnamefont{Perrin}},
  \bibinfo{author}{\bibfnamefont{J.}~\bibnamefont{Tailleur}},
  \bibinfo{author}{\bibfnamefont{G.}~\bibnamefont{Nogues}},
  \bibinfo{author}{\bibfnamefont{M.}~\bibnamefont{Brune}},
  \bibinfo{author}{\bibfnamefont{J.~M.} \bibnamefont{Raimond}},
  \bibnamefont{and} \bibinfo{author}{\bibfnamefont{S.}~\bibnamefont{Haroche}},
  \bibinfo{journal}{Phys. Rev. Lett.} \textbf{\bibinfo{volume}{93}},
  \bibinfo{pages}{103001} (\bibinfo{year}{2004}).

\bibitem[{\citenamefont{Hogan and Merkt}(2008)}]{hogan:043001}
\bibinfo{author}{\bibfnamefont{S.~D.} \bibnamefont{Hogan}} \bibnamefont{and}
  \bibinfo{author}{\bibfnamefont{F.}~\bibnamefont{Merkt}},
  \bibinfo{journal}{Phys. Rev. Lett.} \textbf{\bibinfo{volume}{100}},
  \bibinfo{pages}{043001} (\bibinfo{year}{2008}).

\bibitem[{\citenamefont{Dutta et~al.}(2000)\citenamefont{Dutta, Guest,
  Feldbaum, Walz-Flannigan, and Raithel}}]{dutta00}
\bibinfo{author}{\bibfnamefont{S.~K.} \bibnamefont{Dutta}},
  \bibinfo{author}{\bibfnamefont{J.~R.} \bibnamefont{Guest}},
  \bibinfo{author}{\bibfnamefont{D.}~\bibnamefont{Feldbaum}},
  \bibinfo{author}{\bibfnamefont{A.}~\bibnamefont{Walz-Flannigan}},
  \bibnamefont{and} \bibinfo{author}{\bibfnamefont{G.}~\bibnamefont{Raithel}},
  \bibinfo{journal}{Phys. Rev. Lett.} \textbf{\bibinfo{volume}{85}},
  \bibinfo{pages}{5551} (\bibinfo{year}{2000}).

\bibitem[{\citenamefont{Choi et~al.}(2005{\natexlab{a}})\citenamefont{Choi,
  Guest, Povilus, Hansis, and Raithel}}]{choi:243001}
\bibinfo{author}{\bibfnamefont{J.-H.} \bibnamefont{Choi}},
  \bibinfo{author}{\bibfnamefont{J.~R.} \bibnamefont{Guest}},
  \bibinfo{author}{\bibfnamefont{A.~P.} \bibnamefont{Povilus}},
  \bibinfo{author}{\bibfnamefont{E.}~\bibnamefont{Hansis}}, \bibnamefont{and}
  \bibinfo{author}{\bibfnamefont{G.}~\bibnamefont{Raithel}},
  \bibinfo{journal}{Phys. Rev. Lett.} \textbf{\bibinfo{volume}{95}},
  \bibinfo{pages}{243001} (\bibinfo{year}{2005}{\natexlab{a}}).

\bibitem[{\citenamefont{Choi et~al.}(2005{\natexlab{b}})\citenamefont{Choi,
  Guest, Hansis, Povilus, and Raithel}}]{choi:253005}
\bibinfo{author}{\bibfnamefont{J.-H.} \bibnamefont{Choi}},
  \bibinfo{author}{\bibfnamefont{J.~R.} \bibnamefont{Guest}},
  \bibinfo{author}{\bibfnamefont{E.}~\bibnamefont{Hansis}},
  \bibinfo{author}{\bibfnamefont{A.~P.} \bibnamefont{Povilus}},
  \bibnamefont{and} \bibinfo{author}{\bibfnamefont{G.}~\bibnamefont{Raithel}},
  \bibinfo{journal}{Phys. Rev. Lett.} \textbf{\bibinfo{volume}{95}},
  \bibinfo{pages}{253005} (\bibinfo{year}{2005}{\natexlab{b}}).

\bibitem[{\citenamefont{Gerritsma et~al.}(2007)\citenamefont{Gerritsma,
  Whitlock, Fernholz, Schlatter, Luigjes, Thiele, Goedkoop, and
  Spreeuw}}]{gerritsma:033408}
\bibinfo{author}{\bibfnamefont{R.}~\bibnamefont{Gerritsma}},
  \bibinfo{author}{\bibfnamefont{S.}~\bibnamefont{Whitlock}},
  \bibinfo{author}{\bibfnamefont{T.}~\bibnamefont{Fernholz}},
  \bibinfo{author}{\bibfnamefont{H.}~\bibnamefont{Schlatter}},
  \bibinfo{author}{\bibfnamefont{J.~A.} \bibnamefont{Luigjes}},
  \bibinfo{author}{\bibfnamefont{J.-U.} \bibnamefont{Thiele}},
  \bibinfo{author}{\bibfnamefont{J.~B.} \bibnamefont{Goedkoop}},
  \bibnamefont{and} \bibinfo{author}{\bibfnamefont{R.~J.~C.}
  \bibnamefont{Spreeuw}}, \bibinfo{journal}{Phys. Rev. A}
  \textbf{\bibinfo{volume}{76}}, \bibinfo{pages}{033408}
  (\bibinfo{year}{2007}).

\bibitem[{\citenamefont{Whitlock et~al.}(2009)\citenamefont{Whitlock,
  Gerritsma, Fernholz, and Spreeuw}}]{Whitlock2009}
\bibinfo{author}{\bibfnamefont{S.}~\bibnamefont{Whitlock}},
  \bibinfo{author}{\bibfnamefont{R.}~\bibnamefont{Gerritsma}},
  \bibinfo{author}{\bibfnamefont{T.}~\bibnamefont{Fernholz}}, \bibnamefont{and}
  \bibinfo{author}{\bibfnamefont{R.~J.~C.} \bibnamefont{Spreeuw}},
  \bibinfo{journal}{New J. Phys.} \textbf{\bibinfo{volume}{11}},
  \bibinfo{pages}{023021} (\bibinfo{year}{2009}).

\bibitem[{\citenamefont{Hezel et~al.}(2006)\citenamefont{Hezel, Lesanovsky, and
  Schmelcher}}]{hezel:223001}
\bibinfo{author}{\bibfnamefont{B.}~\bibnamefont{Hezel}},
  \bibinfo{author}{\bibfnamefont{I.}~\bibnamefont{Lesanovsky}},
  \bibnamefont{and}
  \bibinfo{author}{\bibfnamefont{P.}~\bibnamefont{Schmelcher}},
  \bibinfo{journal}{Phys. Rev. Lett.} \textbf{\bibinfo{volume}{97}},
  \bibinfo{pages}{223001} (\bibinfo{year}{2006}).

\bibitem[{\citenamefont{Hezel et~al.}(2007)\citenamefont{Hezel, Lesanovsky, and
  Schmelcher}}]{hezel:053417}
\bibinfo{author}{\bibfnamefont{B.}~\bibnamefont{Hezel}},
  \bibinfo{author}{\bibfnamefont{I.}~\bibnamefont{Lesanovsky}},
  \bibnamefont{and}
  \bibinfo{author}{\bibfnamefont{P.}~\bibnamefont{Schmelcher}},
  \bibinfo{journal}{Phys. Rev. A} \textbf{\bibinfo{volume}{76}},
  \bibinfo{pages}{053417} (\bibinfo{year}{2007}).

\bibitem[{\citenamefont{Mayle et~al.}(2007)\citenamefont{Mayle, Hezel,
  Lesanovsky, and Schmelcher}}]{mayle:113004}
\bibinfo{author}{\bibfnamefont{M.}~\bibnamefont{Mayle}},
  \bibinfo{author}{\bibfnamefont{B.}~\bibnamefont{Hezel}},
  \bibinfo{author}{\bibfnamefont{I.}~\bibnamefont{Lesanovsky}},
  \bibnamefont{and}
  \bibinfo{author}{\bibfnamefont{P.}~\bibnamefont{Schmelcher}},
  \bibinfo{journal}{Phys. Rev. Lett.} \textbf{\bibinfo{volume}{99}},
  \bibinfo{pages}{113004} (\bibinfo{year}{2007}).

\bibitem[{\citenamefont{Mayle et~al.}(2009)\citenamefont{Mayle, Lesanovsky, and
  Schmelcher}}]{mayle:041403}
\bibinfo{author}{\bibfnamefont{M.}~\bibnamefont{Mayle}},
  \bibinfo{author}{\bibfnamefont{I.}~\bibnamefont{Lesanovsky}},
  \bibnamefont{and}
  \bibinfo{author}{\bibfnamefont{P.}~\bibnamefont{Schmelcher}},
  \bibinfo{journal}{Phys. Rev. A} \textbf{\bibinfo{volume}{79}},
  \bibinfo{pages}{041403(R)} (\bibinfo{year}{2009}).

\bibitem[{\citenamefont{Lesanovsky and Schmelcher}(2005)}]{lesanovsky:053001}
\bibinfo{author}{\bibfnamefont{I.}~\bibnamefont{Lesanovsky}} \bibnamefont{and}
  \bibinfo{author}{\bibfnamefont{P.}~\bibnamefont{Schmelcher}},
  \bibinfo{journal}{Phys. Rev. Lett.} \textbf{\bibinfo{volume}{95}},
  \bibinfo{pages}{053001} (\bibinfo{year}{2005}).

\bibitem[{\citenamefont{Marinescu et~al.}(1994)\citenamefont{Marinescu,
  Sadeghpour, and Dalgarno}}]{PhysRevA.49.982}
\bibinfo{author}{\bibfnamefont{M.}~\bibnamefont{Marinescu}},
  \bibinfo{author}{\bibfnamefont{H.~R.} \bibnamefont{Sadeghpour}},
  \bibnamefont{and} \bibinfo{author}{\bibfnamefont{A.}~\bibnamefont{Dalgarno}},
  \bibinfo{journal}{Phys. Rev. A} \textbf{\bibinfo{volume}{49}},
  \bibinfo{pages}{982} (\bibinfo{year}{1994}).

\bibitem[{\citenamefont{Condon and Shortley}(1935)}]{condon35}
\bibinfo{author}{\bibfnamefont{E.~U.} \bibnamefont{Condon}} \bibnamefont{and}
  \bibinfo{author}{\bibfnamefont{G.~H.} \bibnamefont{Shortley}},
  \emph{\bibinfo{title}{The Theory of Atomic Spectra}}
  (\bibinfo{publisher}{Cambridge University Press},
  \bibinfo{address}{Cambridge, England}, \bibinfo{year}{1935}).

\bibitem[{\citenamefont{Pritchard}(1983)}]{PhysRevLett.51.1336}
\bibinfo{author}{\bibfnamefont{D.~E.} \bibnamefont{Pritchard}},
  \bibinfo{journal}{Phys. Rev. Lett.} \textbf{\bibinfo{volume}{51}},
  \bibinfo{pages}{1336} (\bibinfo{year}{1983}).

\bibitem[{\citenamefont{Esslinger et~al.}(1998)\citenamefont{Esslinger, Bloch,
  and H\"ansch}}]{PhysRevA.58.R2664}
\bibinfo{author}{\bibfnamefont{T.}~\bibnamefont{Esslinger}},
  \bibinfo{author}{\bibfnamefont{I.}~\bibnamefont{Bloch}}, \bibnamefont{and}
  \bibinfo{author}{\bibfnamefont{T.~W.} \bibnamefont{H\"ansch}},
  \bibinfo{journal}{Phys. Rev. A} \textbf{\bibinfo{volume}{58}},
  \bibinfo{pages}{R2664} (\bibinfo{year}{1998}).

\bibitem[{\citenamefont{Mewes et~al.}(1996)\citenamefont{Mewes, Andrews, van
  Druten, Kurn, Durfee, and Ketterle}}]{PhysRevLett.77.416}
\bibinfo{author}{\bibfnamefont{M.-O.} \bibnamefont{Mewes}},
  \bibinfo{author}{\bibfnamefont{M.~R.} \bibnamefont{Andrews}},
  \bibinfo{author}{\bibfnamefont{N.~J.} \bibnamefont{van Druten}},
  \bibinfo{author}{\bibfnamefont{D.~M.} \bibnamefont{Kurn}},
  \bibinfo{author}{\bibfnamefont{D.~S.} \bibnamefont{Durfee}},
  \bibnamefont{and} \bibinfo{author}{\bibfnamefont{W.}~\bibnamefont{Ketterle}},
  \bibinfo{journal}{Phys. Rev. Lett.} \textbf{\bibinfo{volume}{77}},
  \bibinfo{pages}{416} (\bibinfo{year}{1996}).

\bibitem[{\citenamefont{Fortagh and Zimmermann}(2007)}]{fortagh:235}
\bibinfo{author}{\bibfnamefont{J.}~\bibnamefont{Fortagh}} \bibnamefont{and}
  \bibinfo{author}{\bibfnamefont{C.}~\bibnamefont{Zimmermann}},
  \bibinfo{journal}{Rev. Mod. Phys.} \textbf{\bibinfo{volume}{79}},
  \bibinfo{pages}{235} (\bibinfo{year}{2007}).

\bibitem[{\citenamefont{Friedrich}(1998)}]{friedrich98}
\bibinfo{author}{\bibfnamefont{H.}~\bibnamefont{Friedrich}},
  \emph{\bibinfo{title}{Theoretical Atomic Physics}}
  (\bibinfo{publisher}{Springer-Verlag}, \bibinfo{address}{Berlin, Germany},
  \bibinfo{year}{1998}), \bibinfo{edition}{2nd} ed.

\bibitem[{\citenamefont{McCurdy et~al.}(2003)\citenamefont{McCurdy, Isaacs,
  Meyer, and Rescigno}}]{PhysRevA.67.042708}
\bibinfo{author}{\bibfnamefont{C.~W.} \bibnamefont{McCurdy}},
  \bibinfo{author}{\bibfnamefont{W.~A.} \bibnamefont{Isaacs}},
  \bibinfo{author}{\bibfnamefont{H.-D.} \bibnamefont{Meyer}}, \bibnamefont{and}
  \bibinfo{author}{\bibfnamefont{T.~N.} \bibnamefont{Rescigno}},
  \bibinfo{journal}{Phys. Rev. A} \textbf{\bibinfo{volume}{67}},
  \bibinfo{pages}{042708} (\bibinfo{year}{2003}).

\bibitem[{\citenamefont{Loew et~al.}()\citenamefont{Loew, Raitzsch, Heidemann,
  Bendkowsky, Butscher, Grabowski, and Pfau}}]{Loew2007}
\bibinfo{author}{\bibfnamefont{R.}~\bibnamefont{Loew}},
  \bibinfo{author}{\bibfnamefont{U.}~\bibnamefont{Raitzsch}},
  \bibinfo{author}{\bibfnamefont{R.}~\bibnamefont{Heidemann}},
  \bibinfo{author}{\bibfnamefont{V.}~\bibnamefont{Bendkowsky}},
  \bibinfo{author}{\bibfnamefont{B.}~\bibnamefont{Butscher}},
  \bibinfo{author}{\bibfnamefont{A.}~\bibnamefont{Grabowski}},
  \bibnamefont{and} \bibinfo{author}{\bibfnamefont{T.}~\bibnamefont{Pfau}},
  \bibinfo{howpublished}{arXiv:0706.2639v1 [quant-ph]}.

\bibitem[{\citenamefont{de~Oliveira et~al.}(2002)\citenamefont{de~Oliveira,
  Mancini, Bagnato, and Marcassa}}]{PhysRevA.65.031401}
\bibinfo{author}{\bibfnamefont{A.~L.} \bibnamefont{de~Oliveira}},
  \bibinfo{author}{\bibfnamefont{M.~W.} \bibnamefont{Mancini}},
  \bibinfo{author}{\bibfnamefont{V.~S.} \bibnamefont{Bagnato}},
  \bibnamefont{and} \bibinfo{author}{\bibfnamefont{L.~G.}
  \bibnamefont{Marcassa}}, \bibinfo{journal}{Phys. Rev. A}
  \textbf{\bibinfo{volume}{65}}, \bibinfo{pages}{031401(R)}
  (\bibinfo{year}{2002}).

\bibitem[{\citenamefont{Folman et~al.}(2002)\citenamefont{Folman, Kr\"uger,
  Schmiedmayer, Denschlag, and Henkel}}]{folman}
\bibinfo{author}{\bibfnamefont{R.}~\bibnamefont{Folman}},
  \bibinfo{author}{\bibfnamefont{P.}~\bibnamefont{Kr\"uger}},
  \bibinfo{author}{\bibfnamefont{J.}~\bibnamefont{Schmiedmayer}},
  \bibinfo{author}{\bibfnamefont{J.}~\bibnamefont{Denschlag}},
  \bibnamefont{and} \bibinfo{author}{\bibfnamefont{C.}~\bibnamefont{Henkel}},
  \bibinfo{journal}{Adv.~At.~Mol.~Opt.~Phys.} \textbf{\bibinfo{volume}{48}},
  \bibinfo{pages}{263} (\bibinfo{year}{2002}).

\bibitem[{\citenamefont{Bunker and Jensen}(1998)}]{bunkerjensen}
\bibinfo{author}{\bibfnamefont{P.~R.} \bibnamefont{Bunker}} \bibnamefont{and}
  \bibinfo{author}{\bibfnamefont{P.}~\bibnamefont{Jensen}},
  \emph{\bibinfo{title}{Molecular Symmetry and Spectroscopy}}
  (\bibinfo{publisher}{NRC Research Press}, \bibinfo{address}{Ottawa, Ontario,
  Canada}, \bibinfo{year}{1998}), \bibinfo{edition}{2nd} ed.

\end{thebibliography}
\end{document}